\documentclass[12pt]{article}

\usepackage{amsmath, amssymb, amsthm}
\newtheorem{prop}{Proposition}

\usepackage{geometry}
\usepackage{setspace}
\usepackage{fancyhdr}
\usepackage{graphicx, float}
\graphicspath{{Figures/}}
\DeclareGraphicsExtensions{.pdf, .png}

\usepackage[justification=centering]{caption}
\usepackage{color}
\definecolor{webgreen}{rgb}{0,.5,0}
\definecolor{webbrown}{rgb}{.6,0,0}
\definecolor{NYUcolor}{rgb}{.6,0,0}
\definecolor{webpurple}{rgb}{0.7,0,0.7}
\definecolor{myred}{RGB}{228, 26, 28}
\definecolor{myblue}{RGB}{55, 126, 184}
\definecolor{mygreen}{RGB}{77, 175, 74}
\definecolor{mypurple}{RGB}{152, 78, 163}
\definecolor{myorange}{RGB}{255, 127, 0}
\definecolor{mygrey}{RGB}{112,128,144}
\definecolor{blue}{rgb}{0,.4,.6}
\definecolor{red}{rgb}{.5,0,0}
\definecolor{green}{rgb}{.1,.5,.1}

\usepackage[hidelinks, 
            breaklinks,
            colorlinks,
            linkcolor={webbrown},
            urlcolor={webpurple}]{hyperref}
            
\hypersetup{
breaklinks = true,
colorlinks=true,
anchorcolor= webbrown,
citecolor= webbrown,
filecolor= webbrown,
linkcolor= webbrown,
menucolor= webbrown,
urlcolor= webbrown,
citebordercolor= 1 0 0,
menubordercolor=1 0 0,
urlbordercolor=1 0 0,
runbordercolor=1 0 0
}

\usepackage{listings, subcaption, forest, tikz-qtree}
\usepackage{babel}
\usepackage{fancyhdr}   
\fancypagestyle{plain}{\pagestyle{fancy}}

\usepackage{lineno}

\usepackage{multicol}
\usepackage{enumitem}

\usepackage{datetime}
\newdateformat{mdy}{\monthname[\THEMONTH] \THEDAY, \THEYEAR}

\usepackage{natbib}
\title{\textbf{The Macroeconomic Effects of  Corporate Tax Reforms}}

\author{\textbf{Francesco Furno}\footnote{Currently: Amazon. Previously: Department of Economics, New York University. Contact: \href{mailto:francesco.furno@nyu.edu}{francesco.furno@nyu.edu}. This paper was completed prior to the author's employment by Amazon. This paper is not sponsored or endorsed by, or associated with Amazon or any of its subsidiaries or affiliates. The views, opinions and positions included in this paper are the author's own and do not reflect the views, opinions and positions of Amazon.\\
I am grateful to my advisors Jaroslav Borovi\v{c}ka, Jess Benhabib and Simon Gilchrist, and to Tim Christensen, Gian Luca Clementi, Francesco Daveri, Francesco Giavazzi, Stefano Rossi, Tom Sargent and B\`{a}lint Sz\H{o}ke for their encouragement. I also thank Mark Gertler, Fran\c{c}ois Gourio, Federico Kochen,  Virgiliu Midrigan, Brandon Pecoraro, Diego Perez, and participants to the NYU Macro Lunch, the BFI's Macro-Financial Modeling Summer Session, the St. Louis Fed, and the Chicago Fed for their comments.}}

\begin{document}
\pagenumbering{gobble}
\maketitle

\vspace{-9mm}
\begin{abstract}
\begin{small}
\noindent Using aggregate, sectoral, and firm-level data, this paper examines the effects of two major U.S. corporate tax cuts. The Tax Cuts and Jobs Act (TCJA-17) led to large shareholder payouts but modest aggregate stimulus, while Kennedy’s 1960s tax cuts stimulated output and investment with minimal payout impact. To explain this divergence, I incorporate tax depreciation policy and a pass-through business sector into a neoclassical growth model. The model suggests that accelerated depreciation and a large pass-through share dampen stimulus from corporate tax rate reductions, and that Kennedy’s cuts boosted output four times more per dollar of lost revenue than the TCJA-17. \end{small} \\

\end{abstract}

\newpage
\pagenumbering{arabic}

\section{Introduction}
This paper investigates the response of the US economy to the recent Tax Cuts and Jobs Act of 2017 (TCJA-17) and Kennedy's corporate tax cuts of the early 1960s, and proposes a framework to think about the macroeconomic effects of such interventions. Both reforms featured a major corporate tax reduction component, but produced different effects on the economy. On the one hand, the TCJA-17 was followed by a large increase in payouts to shareholders, but not by a major stimulus to investment and production. On the other hand, Kennedy's tax cuts of the 1960s were followed by a large increase in output and capital accumulation, but not in payouts to shareholders. Such a duality may appear puzzling. If one thinks of corporate taxes as a form of capital taxation, the lack of stimulus in the aftermath of the TCJA-17 is certainly surprising. If, instead, one thinks of corporate tax cuts as an improductive transfer to shareholders, the large boost to economic activity following Kennedy's cuts is perplexing. 

Understanding the macroeconomic effects of corporate tax reforms has proven to be a challenging task, and a large body of work has tackled this question from different angles (e.g., \cite{barrofurman2018}, \cite{auerbach2018measuring}, \cite{chodorow2024tax}, \cite{gale2024sweeping}, \cite{crump2025large}). Existing macroeconomic frameworks, however, typically abstract from one or both of two institutional features of the tax code that are central to understanding why different corporate tax reforms produce different outcomes: tax depreciation policy and the distinction between c-corporations and pass-through businesses. Once embedded in a standard macroeconomic model and properly calibrated, these two features successfully reproduce the duality of responses documented empirically. Intuitively, this happens because the presence of multiple tax instruments (tax rate, tax depreciation, legal form of organization) produces differential effects on marginal and average business tax rates imposed on the productive sector.

Before presenting the empirical evidence and the theoretical framework, it is useful to briefly review these two institutional elements. Tax depreciation policy governs how fast businesses can deduct the cost of investment from their tax base. When investment is allowed to be deducted over a short period of time, the schedule is said to be `accelerated', and the resulting higher present value of deductions reduces the effective distortion to investment caused by the statutory tax rate.\footnote{For example, see \cite{zwick2017}, \cite{ohrn2018effect}, and \cite{ohrn2019effect}. From a theoretical perspective, the importance of tax depreciation policy is known at least since \cite{smith1963tax} and \cite{hall1967tax}. For an analytical treatment of tax depreciation and some stylized facts, see the second chapter in \cite{furno2022essays}.}

Pass-through businesses are the second institutional element of the analysis. In the US, only c-corporations are subject to corporate income taxation. All other legal forms of organization (s-corporations, partnerships, and sole-proprietorships) are `pass-through', in the sense that their earnings are not subject to firm-level taxation and are `passed through' to their owners. The share of economic activity taking place in pass-through businesses has risen substantially over the last few decades, reaching roughly $40\%$ in 2017, compared to $25\%$ in the early 1960s.\footnote{Several recent contributions have documented some of the implications and issues arising from this pass-through status. For example, see \cite{cooper2016business}, \cite{clarke2017business}, \cite{chen2018corporate}, \cite{smith2019capitalists}, \cite{barro2020taxes}, \cite{kopczuk2020business}, \cite{bhandari2021sweat}, \cite{smith2022rise}, \cite{dyrda2025rise}.} Because a corporate tax cut only applies to c-corporations, the aggregate effect is diluted by the presence of pass-through businesses, which are not only excluded from the cut but also put at a competitive disadvantage relative to the newly-advantaged c-corporate sector.

To illustrate the duality of responses, I present several pieces of empirical evidence. To estimate the aggregate effect of the TCJA-17, I compare pre-reform professional forecasts with actual outcomes for several macroeconomic aggregates, extending the work done by \cite{kopp2019us}. I then adopt the same strategy to analyze the response of c-corporations by aggregating firm-level forecasts from the IBES dataset and firm-level actuals from Compustat. The results indicate a modest response of aggregate output and investment alongside a dramatic loss of corporate tax revenues ($\approx -75\%$ in 2018 excluding revenues from repatriated earnings), while c-corporations exhibited a much larger response of investment and a very large increase in payouts to shareholders. I confirm that these patterns are driven by the corporate tax provisions by documenting a reallocation of economic activity from pass-through businesses to c-corporations using publicly available IRS data, and by providing cross-sectional evidence on the mechanism following the identification strategy in \cite{zwick2017}.

I then gather empirical evidence on Kennedy's tax cuts. The data is significantly sparser for this historical period, but it nonetheless suggests a large increase in investment and output driven both in the aggregate and for c-corporations, and a lack of increase in payouts to shareholders.

To rationalize these empirical results, I propose a theoretical mechanism that leverages tax depreciation policy and pass-through businesses and that can be embedded in a framework as simple as a neoclassical growth model with two sectors: a c-corporate sector subject to corporate taxation with an explicit tax depreciation schedule, and a pass-through sector not subject to corporate taxation. 

In line with a canonical neoclassical model of firm investment, corporate tax changes affect the economy primarily through the investment decision of c-corporations, which depends not only on the tax rate but also on tax depreciation policy. The possibility to deduct investment from the tax base partially counteracts the distortion introduced by the tax rate: the faster investment is deducted, the smaller the distortion to the rate of return on investment. As a result, when tax depreciation policy is very accelerated --- as it was in 2017 --- a rate reduction does little to stimulate investment. When tax depreciation is not accelerated --- as in the early 1960s --- a rate cut provides a substantial boost.

However, as this is one of the key take-aways of this paper, a corporate tax reform can affect the marginal tax rate and the average tax rate differently. A statutory rate cut always reduces the tax bill (i.e., lowers the average tax rate), but the resulting tax savings flow to investment only when the reform materially increases the rate of return (i.e., when it decreases the marginal tax rate). When the marginal tax rate does not decrease--because pre-reform depreciation policy was already accelerated--the savings are distributed to shareholders instead. This distinction, combined with the dilution caused by the pass-through sector, explains both the modest aggregate stimulus and the large shareholder payouts observed after the TCJA-17, as well as the strong investment response and absence of payout growth after Kennedy's cuts. To quantify this duality, I compute the model-implied corporate tax multiplier for each tax reform and show that, for every dollar of lost corporate tax revenues, Kennedy's tax cuts stimulated output roughly four times more than the TCJA-17, and that a large part of this difference can be attributed to differences in pre-reform tax depreciation policy. 

The results in this paper may appear at odds with the common approach of treating corporate taxation as a form of capital taxation. I provide a formal mapping between corporate taxes and capital taxes showing that they are distinct instruments, and that both tax depreciation policy and pass-through businesses drive a wedge between them. I also show that standard macroeconomic approaches to modeling corporate taxes would fail to rationalize the empirical evidence presented here. Finally, I characterize the distortions introduced by US corporate tax policy over the last few decades. This last exercise shows that US policy-makers have by now removed most of the distortions introduced by corporate taxes, but this also implies that they are running out of ammunition: further reductions of corporate tax rates are unlikely to provide strong stimulus to the economy.

\subsection{Relation to the Literature}
The main contribution of this paper is to document a duality of responses of the US economy to two  major corporate tax reductions, and to propose a unified framework able to rationalize the joint response of a wide set of variables across these two episodes.\footnote{There are several policy papers that provide forecasts for the effect of major corporate tax reforms, but they are usually limited to the response of output and corporate tax revenues and do not quantify the importance of tax depreciation policy and pass-through businesses in shaping the effect of each reform.} This is achieved by explicitly analyzing tax depreciation policy and pass-through businesses while keeping the economic environment as simple as possible, in order to illustrate the mechanisms at play transparently and to showcase the explanatory power of these two elements of the tax code.

Relative to the existing macroeconomic literature, the main contribution of this paper is to illustrate the qualitative and quantitative implications of considering both pass-through businesses and tax depreciation policy when studying the macroeconomic effects of a corporate tax reform.\footnote{Papers in macroeconomics that consider tax depreciation policy explicitly include \cite{mertensravn2011}, \cite{winberry2021lumpy}, and \cite{baley2022macroeconomics}. \cite{sedlacek2019reviving} consider full expensing of investment to study the TCJA-17, but otherwise restrict tax depreciation to economic depreciation. Papers that consider pass-through businesses include \cite{chen2018corporate}, \cite{bhandari2021sweat}, and \cite{zeida2021tax}. Examples of papers that abstract from both include \cite{conesa2013}, \cite{erosa2019}, and \cite{chari2020optimal}. \cite{acemoglu2020does} calibrate a capital tax to the effective tax rate paid by corporations, which recovers the tax wedge on the firm's Euler Equation but fails to measure cash-flows correctly.}

The most closely related work is \cite{chodorow2024tax}, who estimate firm-level investment elasticities to various provisions of the TCJA-17 using administrative tax return data, and use a structural model to aggregate these responses and conduct counterfactuals. Their findings are broadly complementary: they document that accelerated depreciation generates more investment per dollar of revenue lost than rate cuts, and their estimated aggregate investment response is quantitatively consistent with the model predictions presented here. The main distinction is one of scope and approach: while their focus is on estimating investment responses to the TCJA-17, this paper proposes a general equilibrium framework that jointly explains a wider set of variables --- including payouts to shareholders --- across two historical episodes, and identifies tax depreciation policy and the pass-through sector as the key drivers of the divergence.

While the intuition that the effect of a corporate tax rate cut on corporate investment depends on tax depreciation policy has been theorized by the public finance literature at least since \cite{hall1967tax}, this insight in isolation is not sufficient to pin down the macroeconomic response of the economy to a corporate tax reform. A general equilibrium modeling approach, instead,  naturally provides restrictions on the comovement of a large set of variables of interest.

The paper also provides some novel empirical results. Following the TCJA-17, this paper documents a small impact on aggregate variables by extending and complementing \cite{kopp2019us}, who focus on investment. It also construct a novel measure of overall c-corporate activity using firm-level data,and leverages publicly available IRS data - some of which have been manually digitalized - to document a shift of economic activity from pass-through businesses to c-corporations after a corporate tax reduction.

Finally, this paper provides two analytic results related to the capital taxation literature in macroeconomics: a formal mapping showing that tax depreciation policy and pass-through businesses drive a wedge between corporate and capital taxes --- implying that thinking of corporate taxes as capital taxation, as in \cite{lucas1990supply} and \cite{mankiw2009optimal}, can be misleading --- and a result showing that the corporate tax can efficiently collect positive revenues in the long run even under full expensing, echoing \cite{abel2007optimal}.

\section{Empirical Evidence on the TCJA-17}
\label{TCJA17}
This section presents empirical evidence on the effects of the TCJA-17. I first examine the response of macroeconomic and c-corporate variables by comparing their actual paths with pre-reform professional forecasts. Since activity of the c-corporate sector is not officially measured, I aggregate firm-level data and forecasts from Compustat and IBES to obtain sectoral aggregates. The results indicate a modest response of macroeconomic aggregates, but a more pronounced response in the c-corporate sector.

Using IRS publicly available data, I then document a  shift in activity from pass-through businesses toward c-corporations after the reform, consistently with the idea that the corporate tax rate provisions of the TCJA-17 provided greater incentives to c-corporations. Finally, to provide additional evidence on the mechanism, I leverage the cross-section of c-corporations in Compustat. Specifically, I estimate the differential investment response to changes to the ``corporate tax wedge'', which measures the effective marginal tax rate on new investment. The results show that reductions in this wedge are positively associated with higher investment levels.

\subsection{The TCJA-17: Corporate Provisions}
\label{TCJA17_Provisions}

It is common for major US tax reforms to include provisions affecting a variety of tax instruments, and the Tax Cuts and Jobs Act of 2017 is no exception. For example, the reform included changes to individual income taxation, to the estate tax exemption, to the individual mandate penalty, to international tax rules, and introduced a deduction for pass-through income. Since corporate taxation is the main focus of this paper, I focus on the main corporate provisions of the TCJA-17, summarized in \autoref{Table_TCJA17_Provisions}.

The TCJA-17 was signed into law on December 22 2017, and the vast majority of its provisions became effective in January 2018. Its two main corporate provisions are a permanent cut to the statutory corporate tax rate from $35\%$ to $21\%$ and a temporary five-year increase in bonus depreciation followed by a phase-out period for assets with an estimated life less than 20 years - i.e.  fixed  capital asset that are not structures. Another important provision reduces the ability of businesses to deduct interest payments on debt from their tax base, while the remaining provisions are aimed at re-organizing the tax code in an overall revenue-neutral fashion.\footnote{Bonus depreciation, together with the newly-introduced pass-through income deduction for the individual income tax, affect pass-through businesses as well. \cite{goodman2021business} document almost no response of pass-through businesses to the pass-through income deduction, and pass-through businesses tend to be less capital-intensive then c-corporations.}

\begin{table}[H]
\begin{small}
\begin{center}
\caption{Corporate Tax Provisions in the TCJA-17}
\begin{tabular}{lc}
\hline \hline
\multicolumn{1}{c}{\textbf{Provision}}                  & \textbf{\begin{tabular}[c]{@{}c@{}}Static Revenue Change (\$bln) \\ 2018-2020\end{tabular}} \\ \hline
Corporate Tax Rate  from 35\% to 21\%                        & $-357.1$                                                                          \\
Bonus Depreciation Allowance from 50\% to 100\%                                      & $-93.6$                                                                          \\
Interest-Deduction Cap                            & $+45.8$                                                                           \\
Small Business Reform (e.g. Section 179)                & $-34.6$                                                                           \\
Additional Changes to Deductions                        & $+35.9$                                                                           \\
Changes to Loss Treatment                               & $+27.5$                                                                           \\
AMT Repeal                                              & $-20.3$                                                                           \\
Changes for Insurances, Banks and Fin Instruments & $+16.7$                                                                           \\
Changes to Business Credits                             & $+2.1$                                                                           \\
Changes Accounting Methods                              & $+5.6$                                                                            \\ \hline \hline
\end{tabular}
\caption*{\footnotesize \textbf{Source}: JCT Conference Report for H.R.1. \\\scriptsize \textbf{Notes}: The numbers reported in the table are estimated using a ``marginal'' approach. The JCT estimates the effect of each provision by adding one after the other. So, for instance, the change in revenues due to the corporate tax rate reduction is estimated conditional on the repeal of the alternative minimum tax (ATM). As a result, the numbers above should be interpreted carefully due to interactions between different tax provisions.}
\label{Table_TCJA17_Provisions}
\end{center}
\end{small}
\end{table}

\subsection{Aggregate and C-Corporate Response}
\label{TCJA17_Macro_Corps}
To estimate the response of the US economy to the TCJA-17, I compare actual realizations of macroeconomic and c-corporate variables with pre-reform professional forecasts, and interpret the difference as the estimated effect of the reform.\footnote{The idea behind this exercise is the same as in \cite{kopp2019us} whose focus is on aggregate business investment.} More formally, I assume that the future value of a generic variable $y$ is given by:
\[ y_{t+h} = \mathbb{E}[y_{t+h} | \mathcal{I}_t^-]  + \beta_h \cdot \Delta \tau_{t+j}  + \varepsilon_{t+h} \]
where $\mathcal{I}_t^-$ is the information set of the forecasters at time $t$ that excludes corporate tax rates, $ \mathbb{E}[y_{t+h} | \mathcal{I}_t^-] $ is the forecast of $y$ based on such an information set, $\Delta \tau_{t+j}$ is the change in corporate tax rates at time $t+j$ where $j \leq h$, $\beta_h$ is the impact of a corporate tax rate change on the variable of interest, and $\varepsilon_{t+h}$ is the effect of all other unanticipated shocks to the economy. I assume that professional forecasts are unbiased, so that $\mathbb{E}[\varepsilon_{t+h}|\mathcal{I}_t^-] = 0$. I assume that changes to the corporate tax rate are unanticipated. Under these assumptions, the effect of a corporate tax change on the variable of interest is given by:
\[  \beta_h \cdot \Delta \tau_{t+j} = y_{t+h} - \mathbb{E}[y_{t+h} | \mathcal{I}_t^-]  -  \varepsilon_{t+h} \]
and can be estimated by comparing actual realizations of $y$ with pre-reform forecasts. Moreover, uncertainty can be estimated leveraging historical forecast errors.\footnote{I estimate uncertainty by fitting a non-parametric density using a normal kernel and choosing bandwidth according to the Silverman's rule of thumb. The forecast errors are computed over the period 2011-2018 for the macro aggregates (SPF and CBO), and over the period 2011-2017 for C-corporations (IBES). Shaded bands in the rest of the section reflect $68\%$ confidence bands.} Professional forecasts for both macroeconomic aggregates and c-corporations are taken from the summer of 2017, the vintage least likely to incorporate anticipation effects, as I show in \autoref{Appendix_Empirics_TCJA17_Anticipation}.

\subsubsection{Response of Macroeconomic Aggregates}
\label{TCJA17_Macro_Response}

\begin{figure}[H]
\centerline{\includegraphics[scale=0.7]{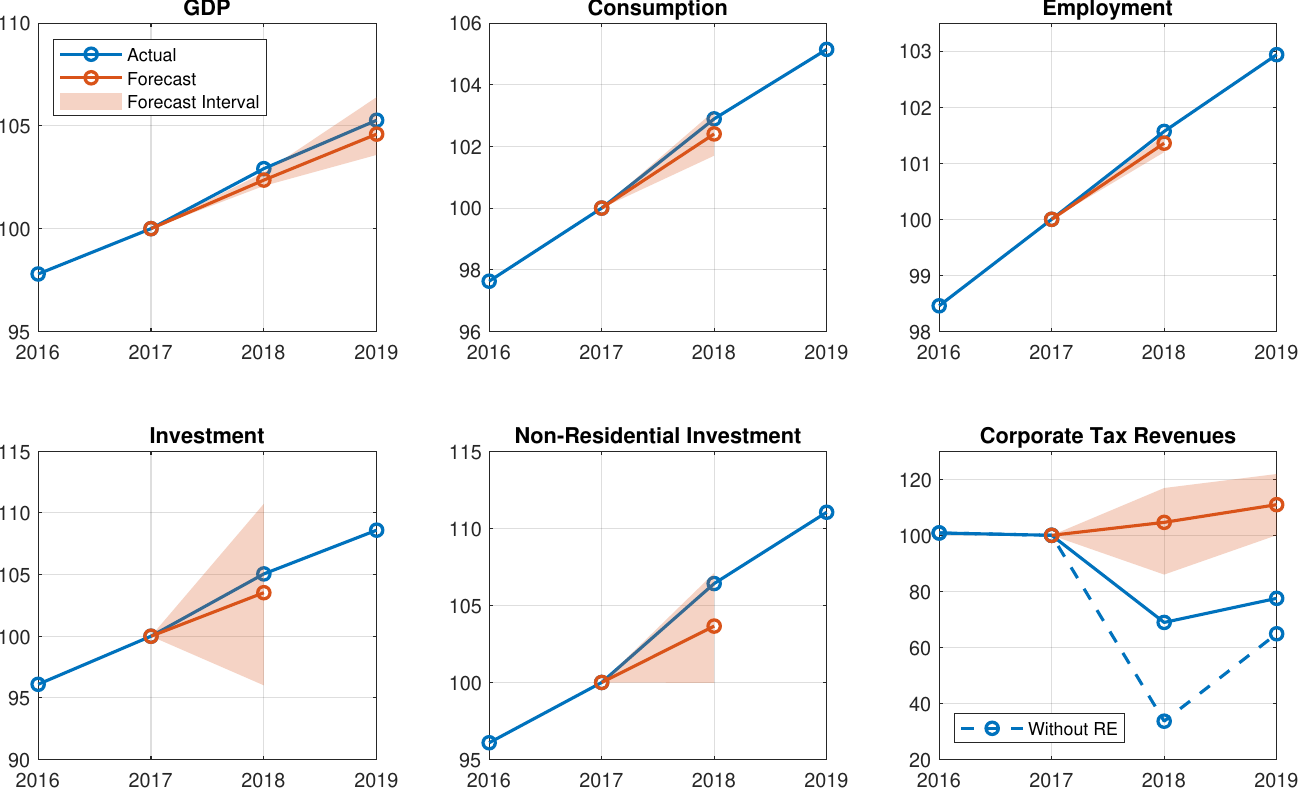}}
\caption{Response of Macroeconomic Aggregates to the TCJA-17 \\\scriptsize \textbf{Notes}: GDP, consumption, investment and non-residential investment are in real terms.  ``Forecast'' refers to the median forecast in the SPF, and the point forecast made by the CBO. The series ``Without RE'' shows corporate tax revenues adjusted to remove the effect of earnings repatriation - the details of the adjustment procedure are in \autoref{Appendix_Empirics_Repatriation}. All values are normalized to $100$ in 2017.}
\label{Figure_TCJA17_Macro}
\end{figure}

Forecasts for macroeconomic aggregates come from the Survey of Professional Forecasters (SPF) and are compared to their NIPA counterparts - except for corporate tax revenues where both actuals and forecasts come from the \text{Update to the Budget and Economic Outlook} produced by the Congressional Budget Office (CBO). The results are reported in \autoref{Figure_TCJA17_Macro}.

The figure shows a modest response of output, consumption, employment, and investment, which is not clearly distinguishable from forecast uncertainty. Interestingly, however, the response of non-residential investment appears larger than that of investment, which is consistent with the idea that the macroeconomic response is driven by the investment decision of the productive sector.

The loss of corporate tax revenue is, instead, dramatic. This is especially true when I perform an adjustment to filter out the effect of earnings repatriation by multinational companies--see \autoref{Appendix_Empirics_Repatriation} for additional details. Since the theoretical framework in this paper features a closed-economy and abstracts from cross-border operations, this adjusted response of corporate revenue becomes a useful  empirical counterpart to assess the predictive power of the theory.

\subsubsection{Response of C-Corporations}
\label{TCJA17_Corporate_Response}
Next, I examine the response of the c-corporate sector as a whole. Surprisingly, actuals and forecasts for the entire c-corporate sector are not produced by neither the government nor the private sector. I thus leverage firm-level data from IBES and Compustat to construct measures of economic activity in the c-corporate sector by aggregating across firms. The IBES database contains actuals and professional forecasts for larger c-corporations, while Compustat contains actuals for all c-corporations subject to public disclosure.

\begin{figure}[H]
\centerline{\includegraphics[scale=0.7]{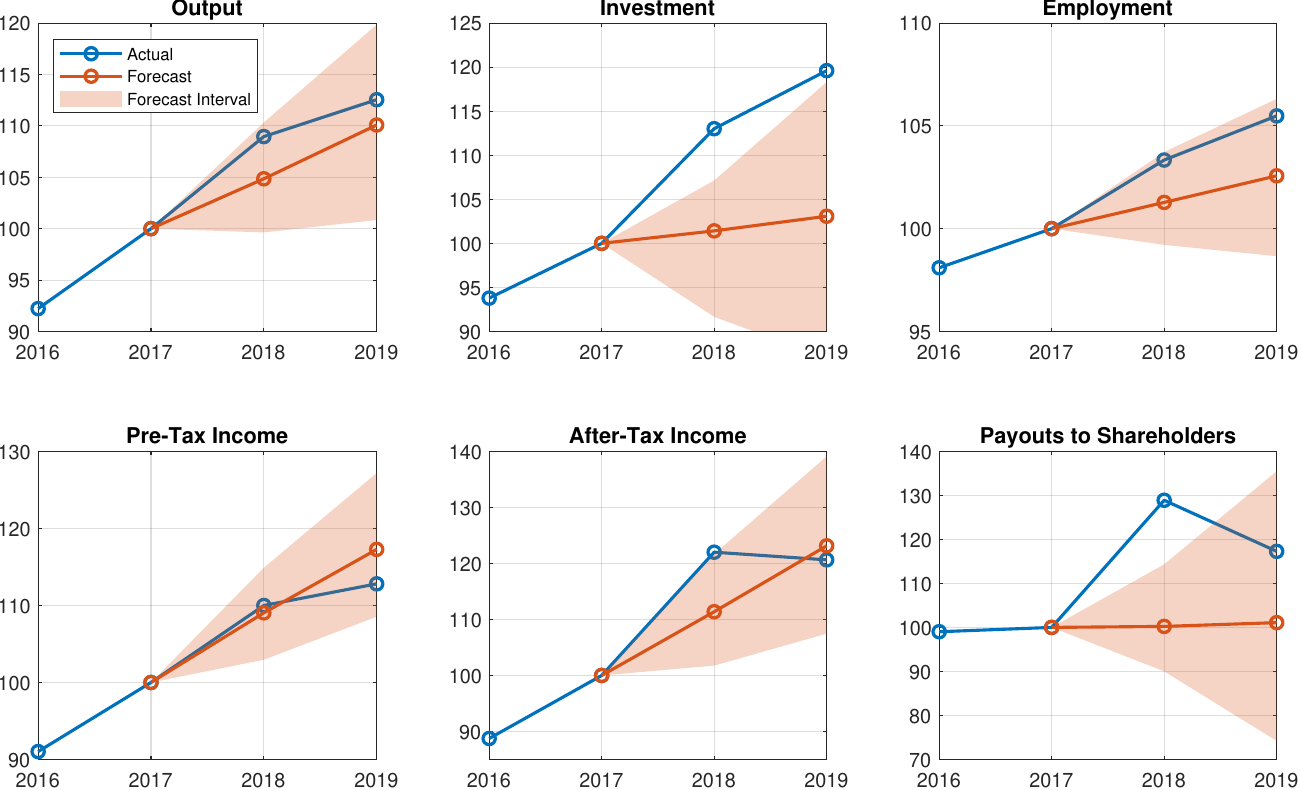}}
\caption{Response of C-Corporations to the TCJA-17 \\ \scriptsize  \textbf{Notes}: Perfectly-balanced panel of $\approx 800$ firms accounting for  $\approx 25\%$ of non-residential investment and $\approx 15\% $ of employment. Data on employment and  share repuchases come from Compustat and their forecasts are constructed by extrapolating their 2-year growth rate. }
\label{Figure_TCJA17_IBES}
\end{figure}

 More formally, I first select a set of firms $\mathcal{F}$ such that both actuals and forecasts from IBES are available for all the years and all the variables considered. This results in a balanced-panel of firms that I can track through the TCJA-17, which in turn  ensures comparability of the c-corporate aggregates across years by removing attrition bias (caused by firms leaving the sample) and selection bias (caused by firms entering the sample). I then compute 
\[  y_{t} = \sum_{f \in \mathcal{F}} y_{f,t}, \quad \hat{y}_{t+h|t} = \sum_{f \in \mathcal{F}} \hat{y}_{f, t+h|t} \]
where $y_{f, t}$ is the value of variable $y$ for firm $f$, and $\hat{y}_{f, t+h|t}$ is the median h-year ahead forecast across professional forecasters for firm $f$. The results are reported in \autoref{Figure_TCJA17_IBES}.

The response of output, investment and employment  appears larger for c-corporations than for the aggregate economy, with actual investment significantly higher than pre-reform forecasts. This is consistent with the idea that the investment decision of c-corporations was affected by the provisions included in the TCJA-17. I also compare the response of pre-tax income, measured by EBITDA, and after-tax income, measured by net income. While pre-tax income in 2018 is in line with forecasts, after-tax income exceeds forecasts because of the reduction in tax-liabilities due to the TCJA-17. Furthermore, the large response of payouts to shareholders, measured as the sum of dividends and share buybacks, suggests that a share of those tax-savings was transferred to shareholders.  The disaggregated results for dividends and share repurchases are reported in \autoref{TCJA17_Payouts_Decomposition}, and show that share repurchases drove payouts in 2018, while dividends adjusted more slowly--in line with theories of dividend smoothing.

\begin{figure}[H]
\centerline{\includegraphics[scale=0.75]{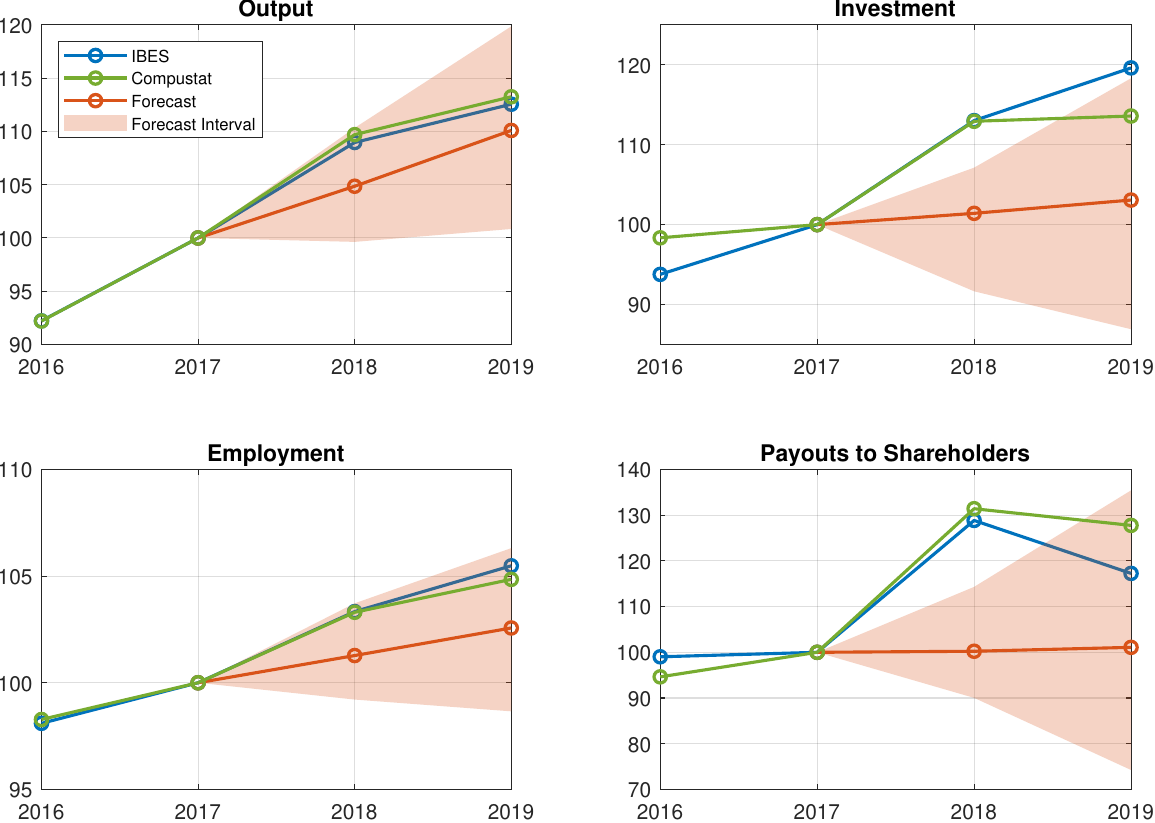}}
\caption{Response of C-Corporations to the TCJA-17: IBES vs Compustat  \\ \scriptsize \textbf{Notes}: IBES comprises a perfectly-balanced panel of $\approx 800$ firms accounting for  $\approx 25\%$ of non-residential investment and $\approx 15\% $ of employment.Compustat comprises a perfectly-balanced panel of $\approx 5000$ firms accounting for  $\approx 50\%$ of non-residential investment and $\approx 30\% $ of employment. }
\label{Figure_TCJA17_IBES_Compustat}
\end{figure}

Since the IBES sample is skewed towards large c-corporations, I compare actuals for firms in the set $\mathcal{F}$ to actuals from  a larger sample of Compustat firms, and the results are reported in \autoref{Figure_TCJA17_IBES_Compustat}. Reassuringly, the response of c-corporations in the  Compustat sample is similar to that in the set $\mathcal{F}$ from the IBES database.  This set covers $25\%$ of aggregate business investment and $15\%$ of aggregate employment, while the Compustat sample covers $50\%$ and $30\%$, respectively. 

\subsection{Reallocation from Pass-Through Businesses to C-Corporations}
\label{TCJA17_Corps_vs_PT}

Businesses in the US can choose to operate under one of four major legal forms of organization: sole-proprietorship, partnership, s-corporation and c-corporation. There are several differences between them, but what matters for this analysis is how each legal form is taxed. The first three forms of organization are pass-through for tax purposes: the business is not taxed directly, but its income is passed through to the owners who are then taxed at the individual income level. C-corporations, instead, are  taxed directly with the corporate income tax.\footnote{Owners of c-corporations are also taxed through the dividend tax once corporate income is distributed, and through the capital-gains tax if they realize a capital gain thanks thanks to a share price increase.}

\begin{figure}[H]
\centerline{\includegraphics[scale=0.675]{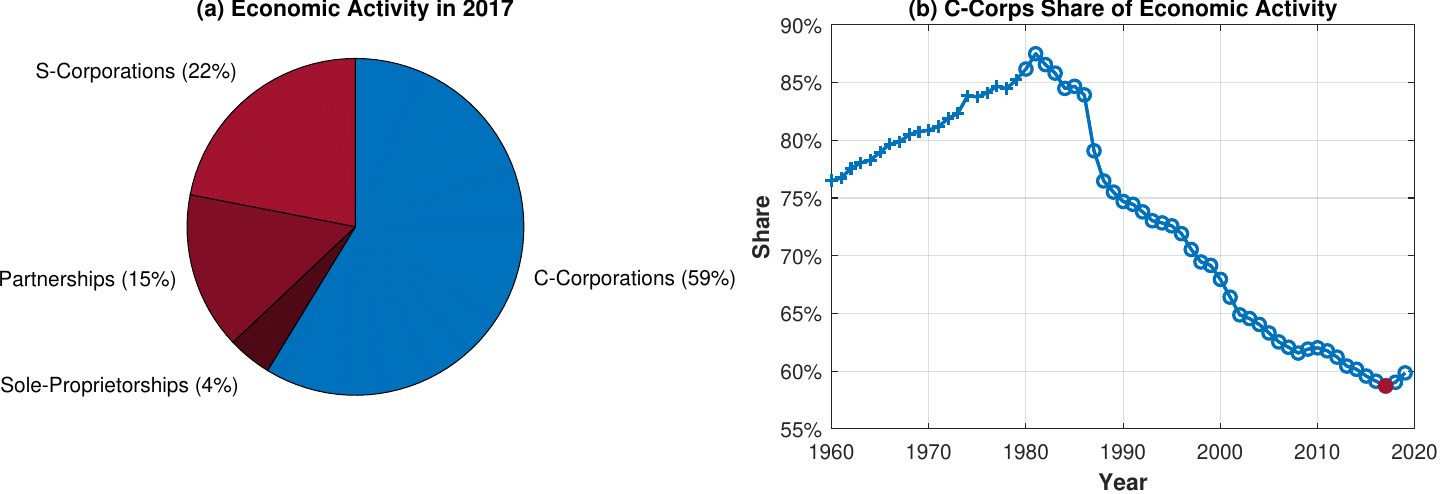}}
\caption{The Size and Evolution of the Pass-Through Sector  \\ \scriptsize  \textbf{Notes}: Economic activity is measured by ``Business Receipts'' from publicly available aggregated tax returns from IRS SOI. Data before 1980 have been manually collected from scanned version of SOI's Business Income Tax Return Reports and Corporation Income Tax Return Reports.}
\label{Figure_TCJA17_PassThrough}
\end{figure}

Panel (a) of
\autoref{Figure_TCJA17_PassThrough} offers a decomposition of US economic activity in 2017 by legal form of organization. In 2017, approximately  $40\%$ of US economic activity was carried out by pass-through businesses and was not subject to corporate income taxation. Also, $25\%$ of economic activity in the corporate sector was not subject to corporate taxation.

Panel (b) of \autoref{Figure_TCJA17_PassThrough} shows the evolution of the share of economic activity that is subject to corporate income taxation since the early 1960s. There are two clear trends. The first one is the steady increase in pass-through economic activity since the tax reforms of the 1980s. The second one  is the rise of c-corporations in the two decades before.\footnote{While the dynamic evolution of the pass-through sector reflects technological, legal and tax considerations, a satisfactory analysis of this phenomenon is  beyond the scope of the analysis.  What this paper emphasizes is that, at any point in time, the aggregate impact of a corporate tax reform depends on the share of economic activity taking place in the  pass-through sector, and this shares has substantially changed over the last decades.}

\begin{figure}[H]
\centerline{\includegraphics[scale=0.7]{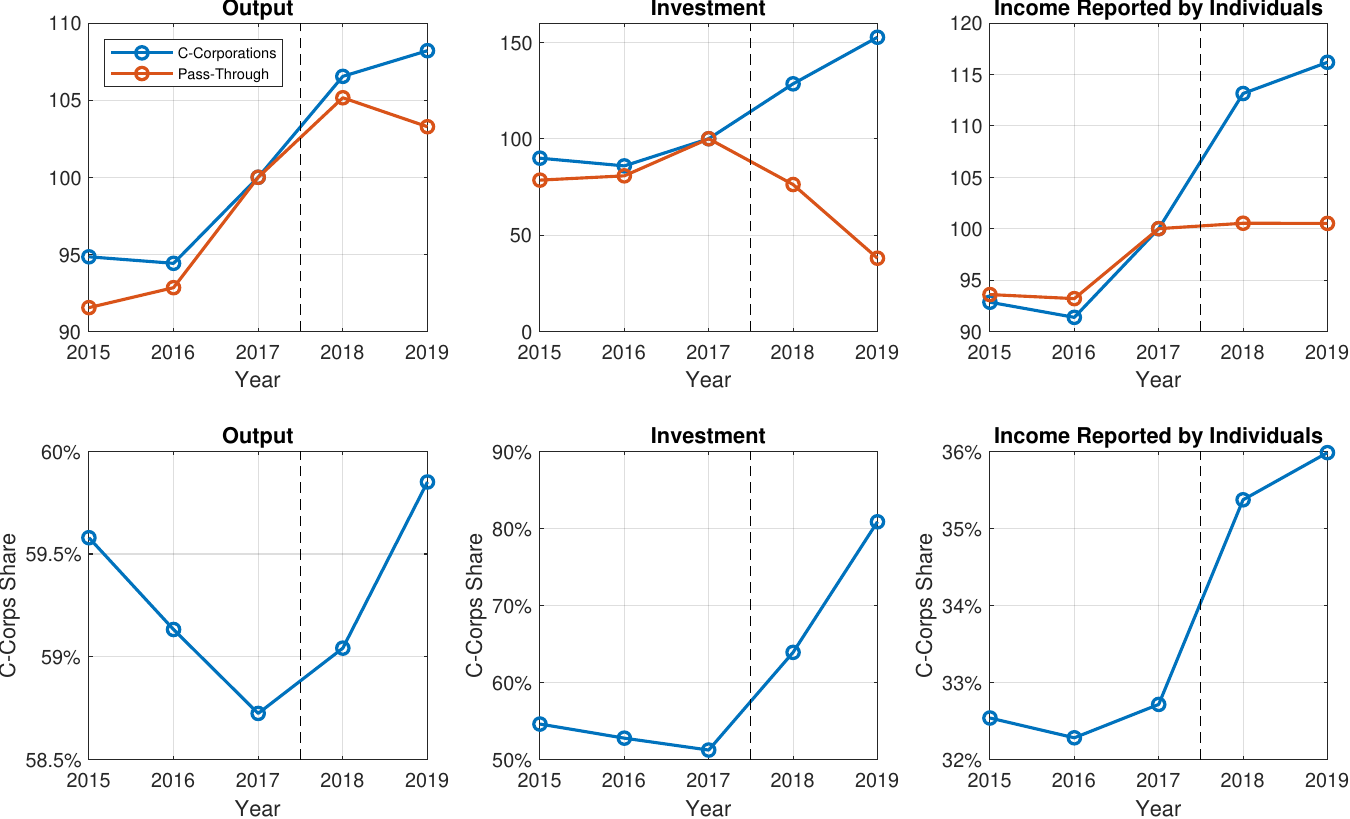}}
\caption{\footnotesize The Shift of Economic Activity from Pass-Through Businesses to C-Corporations \\ \scriptsize \textbf{Notes}: All values are computed from publicly available IRS SOI aggregated tax returns. ``Output'' is measured by ``Business Receipts''. ``Investment'', which is not available for sole-proprietorships, is measured by capital expenditure and is computed as  ``Depreciable Assets'' in year $t$ minus year $t-1$ plus ``Depreciation'' in year $t$. ``Income Reported by Individuals'' defined as the sum of ``Ordinary Dividends'' and ``Qualified Dividends'' for c-corporations, and as the sum of ``Business or Profession Net Income'' and ``Partnership and S-Corporation Net Income'' for pass-through businesses.}
\label{Figure_TCJA17_Corps_vs_PT}
\end{figure}

Given that a material corporate tax reduction should advantage c-corporations relative to pass-through business, I then examine whether the TCJA-17 has resulted in any reallocation across these two sectors. To do so, I leverage publicly available business tax returns from the IRS, and the results are displayed in \autoref{Figure_TCJA17_Corps_vs_PT}.

The top row compares the response of output, investment and income reported by individuals for c-corporations and pass-through businesses, while the bottom row reports the share of c-corporate activity for each of these variables. Tax returns confirm an expansion of the c-corporate sector relative to the pass-through sector following the TCJA-17, and this is especially clear when one looks at the share of activity happening in the c-corporate sector. The  decline  in the years before the reform is consistent with the ``secular rise'' of pass-through businesses (see \cite{smith2019capitalists}, \cite{smith2022rise}, \cite{dyrda2025rise}), but the trend is reversed in 2018 after the TCJA-17.

\subsection{Cross-Sectional Evidence for C-Corporations}
\label{TCJA17_Wedge}
I leverage the cross-section of c-corporations from Compustat and the identification strategy proposed by \cite{zwick2017} to complement the aggregate and sectoral evidence presented so far. I estimate  the differential investment response to differential changes in corporate taxes across 4-digit NAICS industries by estimating the following panel regression:
\begin{gather*}
	\log(i_{t,f,s}) = \alpha_t + \mu_s + \beta \cdot \omega_{t,s} + \delta' X_{t, s, f} + \varepsilon_{t,f,s}
\end{gather*}
where $f$ is the firm index, $X_{t, s, f}$ is a vector of firm-level controls, $\alpha_t$ and $\mu_s$ are fixed effects, and $\omega_{t,s}$ is a ``corporate tax wedge'' given by:
\[   \omega_{t, s} = \frac{1 - \tau^{\pi}_t}{1 - \lambda^{\pi}_{t, s} \tau^{\pi}_t}   \]
where $\tau^{\pi}_t$ is the statutory corporate tax rate in year $t$ and $\lambda^{\pi}_{t, s}$ is the present discounted value of tax depreciation policy in sector $s$ in year $t$. This wedge is a measure of the effective marginal tax rate on new investment and arises both in the traditional cost-of-capital framework of \cite{hall1967tax} and the general equilibrium model introduced later in  \autoref{Theory_Wedge}. The present value of the depreciation schedule is computed as follows:
\[ \lambda^{\pi}_{t, s} = b^{\pi}_t + (1 - b^{\pi}_t)  \lambda^{\pi}_{ s} \]
where $b^{\pi}_t$ is bonus depreciation in year $t$, and $\lambda^{\pi}_{ s}$ is the present discounted value of the representative MACRS tax depreciation schedules for sector $s$ from  \cite{zwick2017}. For each dollar of eligible investment, a fraction $b^{\pi}$ is immediately deducted via bonus depreciation, while the remaining $1-b^{\pi}$ is depreciated according to a tax depreciation schedule. As in \cite{zwick2017}, I ignore structures and focus on fixed assets with an estimated life less than $20$ years (e.g., computers, office furniture, machinery, trucks, etc). The main results are summarized in \autoref{Table_Evidence_Mechanism}.

 \begin{table}[H]
 \begin{small}
 \begin{center}
 \caption{Cross-Sectional Investment Response to the TCJA-17}
 \begin{tabular}{lccccc}
    \hline \hline
  \multicolumn{1}{c}{\textbf{}} & \begin{tabular}[c]{@{}c@{}}(1) \end{tabular}          & \begin{tabular}[c]{@{}c@{}}(2) \end{tabular}           & \begin{tabular}[c]{@{}c@{}}(3) \end{tabular}          & \begin{tabular}[c]{@{}c@{}}(4)\end{tabular}          & \begin{tabular}[c]{@{}c@{}}(5)\end{tabular}          \\ \hline
  \textit{$\omega_{t,s}$}  & \begin{tabular}[c]{@{}c@{}}4.007***\\ (0.434)\end{tabular} & \begin{tabular}[c]{@{}c@{}}8.175***\\ (0.743)\end{tabular}  & \begin{tabular}[c]{@{}c@{}}6.956***\\ (2.023)\end{tabular} & \begin{tabular}[c]{@{}c@{}}6.180***\\ (2.097)\end{tabular} & \begin{tabular}[c]{@{}c@{}}6.180**\\ (2.880)\end{tabular}  \\
  \textit{}                     &                                                            &                                                            &                                                           &                                                           &                                                             \\ \hline
  Firm FE                     & Y                                                          & N                                                          & N                                                         & N                                                         & N                                                         \\
  NAICS FE                       & N                                                          & Y                                                          & Y                                                         & Y                                                         & Y                                                         \\
  Year FE                 & N                                                          & N                                                          & Y                                                         & Y                                                         & Y                                                         \\
  SE Clustering                 & Firm                                                     & NAICS                                                     & NAICS & NAICS                                                    & Firm                                                     \\
  Controls         & N                                                      & N                                                      & N                                                     & Y & Y \\
  Obs                           & 32,802                                                      & 33,551                                                      & 33,551                                                     & 33,190                                                     & 33,190                                                     \\ \hline \hline
  \end{tabular}
 \caption*{\footnotesize \textbf{Notes:} Clustered standard errors in parenthesis. *$p<0.10$, **$p<0.05$, ***$p<0.01$.  Controls include cash, sales, and assets. Sample spans 2014-2020.}
\label{Table_Evidence_Mechanism}
 \end{center}
 \end{small}
 \end{table}

The estimates suggest a robust and precise positive effect of an increase in the corporate tax wedge - i.e. a reduction in the marginal effective tax rate - on investment. The specification with controls, sectoral and time fixed-effects suggests that a (relative) reduction in the marginal effective tax rate produces a (relative) increase in investment by roughly $6\%$.  

In the theoretical counterfactuals presented in \autoref{Theory_Simulating_TCJA17}, I assume that the TCJA-17 reduces the marginal effective corporate tax rate of around $2.8\%$, which multiplied by the  point estimate in columns (4-5) yields an increase in investment by about $17\%$. This magnitude is in line with the evidence on c-corporations presented in \autoref{TCJA17_Corporate_Response}.

\section{Empirical Evidence on Kennedy's Tax Cuts}
Kennedy's tax cuts were legislated and implemented between 1962 and 1965. The Revenue Act of 1962 introduced a $7\%$ investment tax credit for businesses and, in the same year, the IRS also issued a new set of more accelerated tax depreciation guidelines. Both provisions were implemented in 1962. The Revenue Act of 1964 then reduced the top individual tax rate from $91\%$ to $70\%$, reduced individual tax rates across brackets, created the standard deduction, and reduced the corporate tax rate from $52\%$ to $48\%$. The corporate tax rate reduction was implemented in 1964 and 1965. I follow \cite{romer2010} in classifying them as debt-financed.\footnote{\cite{romer2010} also classify these provisions as ``exogenous'', since they were motivated by the desire to increase the long-rung growth rate of the economy. For additional details on Kennedy's tax cuts see \cite{taxfoundation2016}.}

\begin{figure}[H]
\centerline{\includegraphics[scale=0.75]{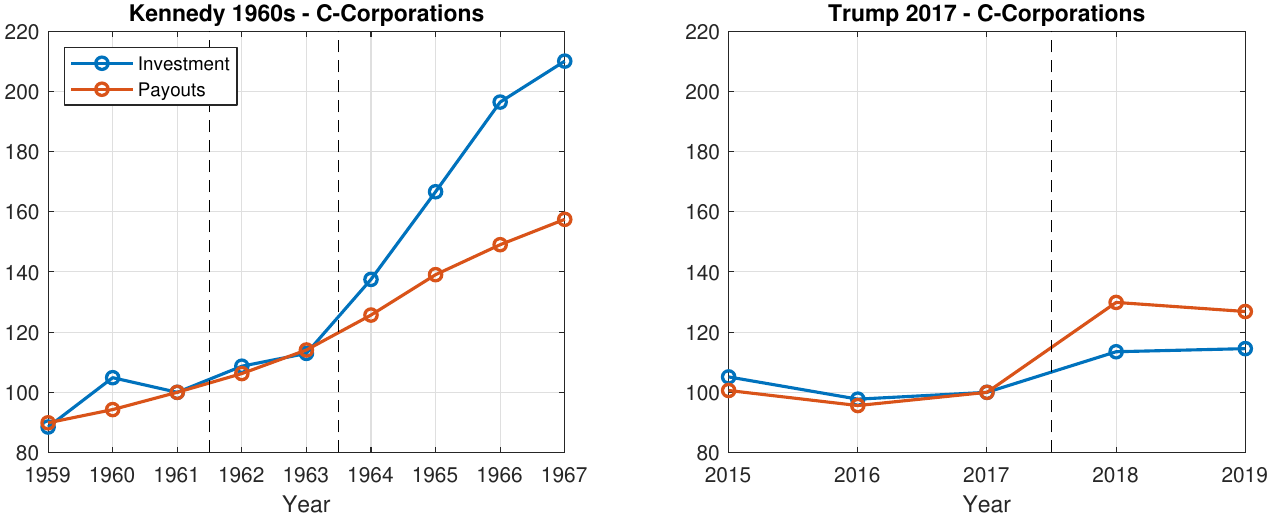}}
\caption{Investment and Payouts for Kennedy's and Trump's Reforms \\ \scriptsize   \textbf{Notes}: Data for c-corporations come from Compustat. For Kennedy's tax cuts, a perfectly-balanced sample of $\approx 600$ c-corporations accounts for $\approx 35 \%$ of business investment. For the TCJA-17, a perfectly-balanced sample of $\approx 4000$ c-corporations accounts for $\approx 40 \%$ of business investment. Share repurchases are included in payouts for TCJA-17, but not for the tax cuts of the 1960s since they were considered a form of market manipulation and largely illegal until 1982. Values are normalized to 100 in 1961 on the left panel  and to 100 in  2017 on the right panel.}
\label{Figure_Kennedy_Empirics}
\end{figure}

Both \cite{hall1967tax} and \cite{cummins1994reconsideration} leverage the cross-section of corporations to estimate the investment response to accelerated depreciation and investment tax credits, and find a material response of investment. Here, I complement their findings and focus on the aggregate response of the economy by leveraging the time series of  investment and payouts to shareholders computed from Compustat and reported in \autoref{Figure_Kennedy_Empirics}.
The increase in payouts to shareholders outweighs the increase in investment after the recent TCJA-17, but not after Kennedy's tax cuts. After the latter, payouts do not appear to deviate much from the existing trend, unlike investment which exhibits a clear acceleration. The increase in capital formation is substantial:  c-corporations' capital expenditure doubled between 1963 and 1967.

\begin{figure}[H]
\centerline{\includegraphics[scale=0.75]{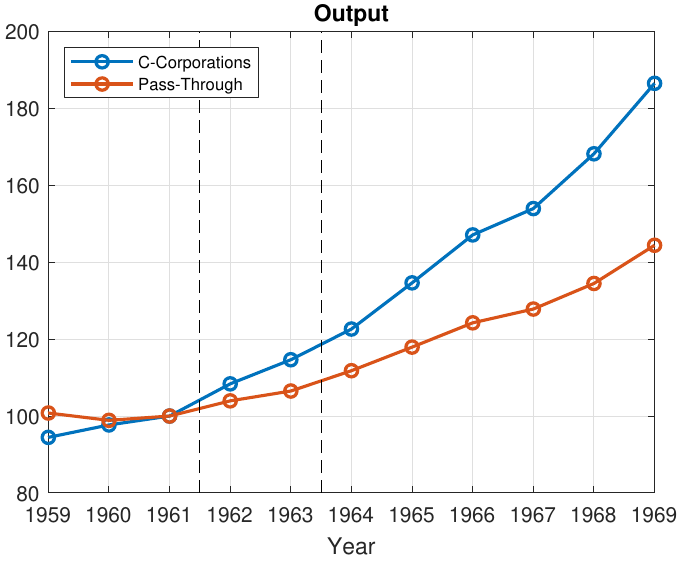}}
\caption{C-Corporations vs Pass-Throughs in 1960 \\ \scriptsize   \textbf{Notes}: All values are computed from the publicly available IRS SOI aggregated tax returns. ``Output'' is measured by ``Business Receipts''. Values are normalized to 100 in 1961.}\label{Figure_Kennedy_Empirics_Reallocation}
\end{figure}

It possible, of course, that the acceleration of economic activity after Kennedy's cuts was not due to the corporate tax provisions. \autoref{Figure_Kennedy_Empirics_Reallocation} mitigates this concern by showing the response of output for c-corporations and pass-through businesses before and after the reform.

The figure displays a stronger acceleration of output for c-corporations relative to pass-through businesses in the years after 1961, consistently with the idea that Kennedy's cuts provided relatively more stimulus to the c-corporate sector.

\section{Theoretical Framework}
This section introduces the theoretical framework and documents its ability to explain the empirical evidence presented so far. To illustrate the proposed mechanism in a clear way, I extend a workhorse neoclassical growth model with pass-through business (modeled as an additional sector), and tax depreciation policy. Despite its simplicity, the model can predict the qualitative response of macroeconomic and c-corporate variables to the TCJA-17 documented in the previous section. I then extend this two-sector neoclassical growth model with endogenous labor supply, a more flexible CES aggregator of consumption across the two sectors, and variable capital utilization, and use this more flexible model to perform policy experiments.

\subsection{A  Model with Tax Depreciation and Pass-Through Firms}
\label{Theory_Baseline_Model}
The model economy is deterministic and populated by a productive sector, a representative household, and a government. The productive sector is further divided into a representative c-corporate sector and a representative pass-through sector. The former is subject to corporate income taxation and distributes its after-tax cash-flows to its shareholders. The latter is not directly subject to taxation, and its cash-flows are `passed-through' to its shareholders. In the rest of the paper, variables relating to the pass-through sector will be denoted with a tilde.

The representative household solves the following optimization problem:
\begin{gather*}
\max_{\{ \hat{c}_t, c_t, \tilde{c}_t, S_{t+1}, \tilde{S}_{t+1}, l_t, \tilde{l}_t  \}} \quad   \sum_{t=0}^{+\infty} \beta^t \frac{\hat{c}_t^{1 - \sigma}}{1 - \sigma}   \\
s.t. \quad 	\hat{c}_t = c_t^{\gamma} \cdot \tilde{c}_t^{1 - \gamma} \\
			c_t + p_t \tilde{c}_t + \Delta S_{t+1} P_t + \Delta \tilde{S}_{t+1} \tilde{P}_t = (1 - \tau^{II}) \cdot \big[ w_t l_t + p_t \tilde{w}_t \tilde{l}_t + S_t d_t + \tilde{S}_t \tilde{d}_t \big] + \text{Transfer}_t   \\
			l_t + \tilde{l}_t = 1, \quad l_t = l, \quad \tilde{l}_t = \tilde{l} \\
 			\Lambda_{t+j,t} \equiv \beta^j \cdot \frac{u'(\hat{c}_{t+j})}{u'(\hat{c}_{t})} \cdot \frac{\partial \hat{c}_{t+j} / \partial c_{t+j}}{\partial \hat{c}_{t} / \partial c_t}
\end{gather*}
where $c_t$ is consumption of goods from c-corporations, $\tilde{c}_t$ is consumption of goods from pass-through businesses, and $\hat{c}_t$ is a consumption bundle constructed using a Cobb-Douglas aggregator.\footnote{C-corporations and pass-through businesses compete even in narrowly-defined industries of the US economy. As an illustrative example, IRS data from 2012 shows  that $52.3\%$ of economic activity in the ``Apparel Manufacturing'' industry (NAICS 315) took place in c-corporations, and the remaining $47.7\%$ in pass-through businesses. Hence the assumption of imperfect substitutability between goods produced by the two sectors.} The good produced by the c-corporate sector is the numeraire, and $p_t$ is the (relative) price of the good produced by pass-through businesses. The household supplies labor inelastically to each sector, and receives wages equal to $w_t$ and $p_t \tilde{w}_t$ each period. She also invests in shares of each sector, that trade at prices $P_t$ and $\tilde{P}_t$.\footnote{In equilibrium, the supply of each type of shares will be fixed and normalized to one.} Ownership of the productive sector entitles the household to dividends $d_t$ from c-corporations, and pass-through income $\tilde{d}_t$ from pass-through businesses. Finally, the household pays individual income taxes and receives transfers from the government. For simplicity, I assume that there is a uniform individual income tax rate $\tau^{II}$ on labor income, dividends and pass-through income.\footnote{In practice, dividends are taxed at a preferential rate, there are numerous deductions and exemptions, and there are tax brackets. Since my main theoretical experiments will involve changing the corporate tax rate while leaving the individual income tax rate unchanged, a uniform individual income tax rate will preserve my main conclusions.} Finally, the household's intertemporal marginal rate of substitution $\Lambda_{t,t+j}$ will be used by the productive sector when making intertemporal decisions.

To better understand how corporate tax reforms affect the economy, I impose as much symmetry as possible between c-corporations and pass-through businesses. Each sector accumulates its own representative capital stock through investment, hires labor competitively, and produces a final good using a constant return-to-scale technology. However, only c-corporations pay corporate income taxes.

\begin{multicols}{2}
\textbf{C-Corporations}
  \begin{align*}
  \max_{\{ d_t, \pi_t, T^{\pi}_t, TB^{\pi}_t, Y_t, l_t, k_{t+1}, i_t \}} \quad &  \sum_{t=0}^{+\infty} \Lambda_{0,t} d_t \\
  s.t. \quad 	d_t &=   \pi_t - T_t^{\pi}  \\
      \pi_t &= Y_t - w_t l_t - i_t\\
      k_{t+1} &= (1 - \delta) k_t + i_t \\
      Y_t &=  k_t^{\alpha} \cdot l_t^{1 - \alpha} \\
      \textcolor{NYUcolor}{T_t^{\pi}} &\textcolor{NYUcolor}{= \tau^{\pi} \cdot TB_t^{\pi}} \\
      \textcolor{NYUcolor}{TB_t^{\pi}} &\textcolor{NYUcolor}{= Y_t - w_t l_t - ID^{\pi}_t}
  \end{align*}

\columnbreak
\textbf{Pass-Through Businesses}
\begin{align*}
\max_{\{ \tilde{d}_t, \tilde{\pi}_t, \tilde{Y}_t, \tilde{l}_t, \tilde{k}_{t+1}, \tilde{i}_t \}} \quad &  \sum_{t=0}^{+\infty} \Lambda_{0,t} \tilde{d}_t \\
s.t. \quad 	\tilde{d}_t &=   \tilde{\pi}_t \\
    \tilde{\pi}_t &= p_t \cdot \Big( \tilde{Y}_t - \tilde{w}_t \tilde{l}_t -  \tilde{i}_t \Big)\\
    \tilde{k}_{t+1} &= (1 - \tilde{\delta}) \tilde{k}_t + \tilde{i}_t \\
    \tilde{Y}_t &=  \tilde{k}_t^{\tilde{\alpha}} \cdot l_t^{1 - \tilde{\alpha}}
\end{align*}
\end{multicols}

Corporate income taxes $T^{\pi}_t$ are computed by multiplying the corporate income tax base $TB_t^{\pi}$ by the statutory corporate income tax rate $\tau^{\pi}$. The corporate income tax base differs from corporate cash-flows because investment is usually not treated as an expense, but is deducted according to a tax depreciation schedule.\footnote{In reality, firms use a mix of capital assets to produce their final goods, and each asset category is potentially subject to a different tax depreciation schedule. Therefore, the capital stock in the model should be interpreted as a representative non-building business capital, and the tax depreciation schedule as a representative tax depreciation schedule for that capital.} As a result, a fraction of present and past investment is deducted from the tax base each period, and this represents the investment deduction $ID^{\pi}_t$ allowed by the tax code.

In general, the investment deduction for a generic period $t$ is given by:
\[ ID^{\pi}_t = \sum_{j=0}^{+\infty} \delta^{\pi}_j \cdot i_{t-j}  \]
where the policy parameters $\{ \delta^{\pi}_j \}_{j=0}^{+\infty}$ represent the percentage of investment from $j$ periods ago that can be deducted from the tax base. Investment is eventually deducted from the tax base in full, so that the policy parameters sum up to one. To improve tractability and build intuition, I approximate the tax depreciation schedule using a declining-balance tax depreciation schedule, which permits the aggregation of all non-depreciated past investment into an auxiliary variable $k^{\pi}_t$.\footnote{\cite{winberry2021lumpy} adopts the same approximation. In \cite{furno2022essays}, I show that the error due to this approximation is negligible in standard economic environments.} The investment deduction can then be rewritten as
\begin{align*}
  \textcolor{black}{ID_t^{\pi}} &= \delta^{\pi} \cdot (i_t + k_t^{\pi}) \\
\text{where} \quad  \textcolor{black}{k_{t+1}^{\pi}}  &= (1 - \delta^{\pi}) \cdot (i_t + k_t^{\pi})
\end{align*}
The auxiliary variable $k_t^{\pi}$ represents the stock of past investment that has not been depreciated for tax purposes yet, and $\delta^{\pi}$ is now the only policy parameter summarizing the tax depreciation schedule, where $\delta^{\pi}_j = \delta^{\pi} \cdot (1 - \delta^{\pi})^j$. In this way, the corporate tax code is fully summarized by the pair $(\tau^{\pi}, \delta^{\pi})$.

To close the model, I introduce a government that collects tax revenues that can go into wasteful spending or into transfers to the representative household:
\begin{align*}
	T_t &= T^{\pi}_t + T^{II}_t \\
	G_t &= \theta \cdot T_t \\
	\text{Transfer}_t &= (1 - \theta) \cdot T_t
\end{align*}
where $T^{II}_t$ are individual income tax revenues, $T_t$ are total tax revenues, and $G_t$ is wasteful spending. The parameter $\theta \in [0,1]$ determines the share of tax revenues that go into wasteful spending. When $\theta=0$, all tax revenues are distributed back to the representative household. Finally, aggregate output and aggregate investment are defined as:
\begin{align*}
  \hat{Y}_t &= Y_t + p_t \tilde{Y}_t \\
  \hat{i}_t &= i_t + p_t \tilde{i}_t.
\end{align*}

\subsubsection{Investment and Tax Liability of C-Corporations}
\label{Theory_Wedge}
In the baseline model, the investment decision of the c-corporate sector is driven by the following Euler Equation:

\[ 1 = \Lambda_{t, t+1} \Bigg[ \textcolor{NYUcolor}{\underbrace{\frac{1 - \lambda^{\pi}_{t+1} \tau^{\pi}}{1 - \lambda^{\pi}_t \tau^{\pi}}}_{\approx 1}} \cdot  (1 - \delta)  + \textcolor{NYUcolor}{\underbrace{\frac{1 - \tau^{\pi}}{1 - \lambda^{\pi}_t \tau^{\pi}}}_{\text{``Corporate Tax Wedge''}}} \cdot  MPK_{t+1}  \Bigg] \]

\[  \text{where} \quad  \textcolor{NYUcolor}{ \lambda^{\pi}_t =  \underbrace{\sum_{j=0}^{+\infty}  \Lambda_{t, t+j} \cdot \big[ (1 - \delta^{\pi})^j \cdot \delta^{\pi} \big]}_{\text{PDV of tax depreciation schedule}} } \]

The distortion to the investment decision introduced by the corporate tax code shows up in the form of a wedge, that I label as the ``corporate tax wedge''. This wedge is jointly determined by the statutory tax rate and the present discounted value of the tax depreciation schedule. This result mirrors  \cite{hall1967tax}, and can be thought of as an extension to general equilibrium thereof.\footnote{This happens because the baseline model is a neoclassical model. In general, when the economic environment is enriched with frictions, it is not possible to summarize the distortions to the investment decision in such a clear-cut way.}

A higher value of $\delta^{\pi}$ reflects a more accelerated tax depreciation policy, which in turn implies that both  $\lambda_t^{\pi}$ and the corporate tax wedge are closer to one. As a result, even when the statutory tax rate is high, the distortions to the investment decision can be small if tax depreciation policy is highly accelerated.

The tax rate and tax depreciation policy also determine the tax bill of c-corporations:
\begin{align*}
   T_t^{\pi} &= \textcolor{NYUcolor}{\tau^{\pi}} \cdot \Big[Y_t - w_t l_t - \sum_{j=0}^{+\infty}  \textcolor{NYUcolor}{\delta^{\pi} \cdot (1 - \delta^{\pi})^j} \cdot i_{t-j}  \Big]
\end{align*}
However, changes to the corporate tax code do not affect the investment decision and the tax bill in the same way. It is possible - and this is key to understand the TCJA-17 -  to conceive a corporate tax reform that leaves the corporate tax wedge almost unchanged, while producing a big change to the corporate tax bill.

\subsubsection{Calibration to the US Economy before the TCJA-17}
\label{Theory_Calibration}

I calibrate the model to the US economy in 2017, just before the TCJA-17. Several parameters - such as the discount rate, the household's IES, economic depreciation and the labor share - are standard. I calibrate labor supply and the exponents of the Cobb-Douglas consumption aggregator to match the relative size of the c-corporate and pass-through sectors.

\begin{table}[H]
\begin{center}
\caption{Calibration of the Baseline Model}
\begin{tabular}{ccc}
  \hline \hline
  \textbf{Parameter} & \textbf{Value} & \textbf{Notes}             \\ \hline
  $\beta$            & 0.94           & Rate of time preferences   \\
  $\sigma$            & 1         & IES \\
  $\delta = \tilde{\delta}$           & 0.10            & Physical depreciation rate \\
  $\alpha = \tilde{\alpha}$           & 0.35           & Labor share   ($=0.65$)              \\
  $l$  & 0.575& C-Corps share of salaries and wages \\
  $\gamma$ & 0.575& C-Corps share of business receipts  \\ \hline
  $\tau^{\pi}$       & 0.35           & Statutory Corporate Tax Rate               \\
  $\delta^{\pi}$     & 0.4823           & Tax Depreciation Rate       \\
  $\tau^{II}$       & 0.135            & Average effective tax rate                             \\
  $\theta$ & 0 & Mimic a debt-financed tax cut \\
 \hline \hline
\end{tabular}
\label{Table_Calibration}
\end{center}
\end{table}

The tax code is calibrated as follows. The corporate tax rate is set equal to the statutory corporate tax rate. The tax depreciation rate $\delta^{\pi}$ is set in such a way that it matches the present discounted value of a representative tax depreciation schedule computed using the same methodology proposed in \cite{zwick2017}. This present discounted value averages tax depreciation schedules for different types of capital assets, and includes the $50\%$ bonus depreciation that was in place in 2017 - see \autoref{Appendix_Calibration_Depreciation} for the details. The individual income tax rate is set equal to the average effective tax rate computed from publicly available individual income tax returns from the IRS. Finally, in order to mimic a debt-financed tax cut, I assume that all tax revenues are transferred back to the representative household by setting $\theta = 0$.

\autoref{Table_Calibration_Fit} shows that the calibrated model's deterministic steady-state is able to reproduce four important empirical moments: corporate profits, dividends, corporate tax revenues and individual income tax revenues as a share of GDP. These moments are not explicitly targeted by the calibration, but the model can match them well because the way the variables are defined in the model is a good approximation of what happens in practice.

\begin{table}[H]
\begin{center}
\caption{Fit of Key Untargeted Moments}
\begin{tabular}{ccc}
\hline \hline
\textbf{Moment} & \textbf{Model (SS)} & \textbf{Data}             \\ \hline
    $\pi / Y$			& 0.08 & 0.10  \\
    $ d /Y$      	 & 0.05 & 0.05  \\
    $ T^{\pi} / Y$ & 0.03 &  0.02 \\
    $ T^{II} / Y$    & 0.10 & 0.08   \\
     \hline \hline
\end{tabular}
\label{Table_Calibration_Fit}
\end{center}
\caption*{\scriptsize \textbf{Notes}: Model (SS) refers to the deterministic steady-state of the model. Data comes from NIPA and span the period 2012-2017. Corporate profit and dividends in the NIPA refer to both c-corporations and s-corporations, thus slightly over-estimating the value for c-corporations alone.}
\end{table}

 Matching these four untargeted moments ensures that the size of the corporate sector and of the government's tax collection in the model is representative of the US economy before the TCJA-17.

\subsection{The TCJA-17: Model vs Data}
\label{Theory_Simulating_TCJA17}
The Tax Cuts and Jobs Act of 2017 is simulated by starting from the calibration in \autoref{Table_Calibration} and introducing an unanticipated permanent change to the following policy parameters:
\begin{itemize}[noitemsep]
  \item A permanent reduction in the corporate tax rate $\tau^{\pi}$ from $35\%$ to $21\%$.
  \item A permanent increase in the tax depreciation rate  $\delta^{\pi}$ from $0.4823$ to $0.8305$.
\end{itemize}
The change to the tax depreciation rate increases the  present discounted value of the representative tax depreciation schedule in steady-state from $\approx 0.94$ to $\approx 0.99$.\footnote{I allow for $90\%$ bonus depreciation, instead of $100\%$, to take into account the fact that the TCJA-17 placed some restrictions on asset eligibility - see \autoref{Appendix_Calibration_Depreciation} for the details. The main results are almost unchanged if I assume $100\%$ bonus depreciation instead.}
While the TCJA-17 increased  bonus depreciation only temporarily,  US policy-makers have repeatedly extended expiring bonus depreciation over the last couple of decades. It is not unreasonable to believe that bonus depreciation will be extended upon expiration, which justifies the assumption of a permanent change. Importantly, since the increase in bonus depreciation  only applies to new investment,  I introduce auxiliary variables to distinguish between old and new investment for tax purposes - see \autoref{Appendix_New_Investment} for details.

\begin{figure}[H]
\centerline{\includegraphics[scale=0.675]{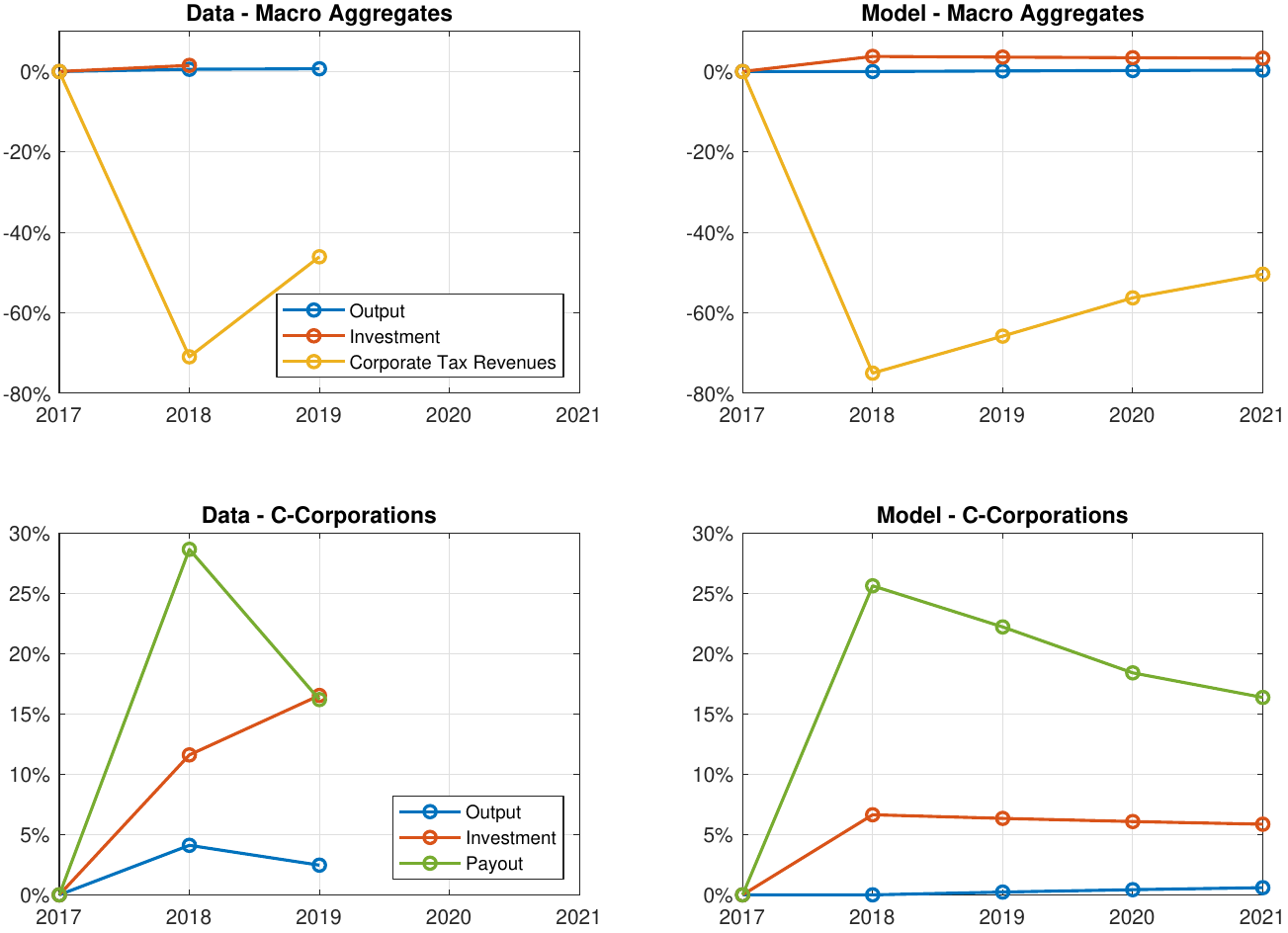}}
\caption{\footnotesize The TCJA-17: Model vs Data \\ \scriptsize \textbf{Notes}: Empirical moments are computed as the difference between the actual realizations and the pre-reform forecasts from \autoref{TCJA17}. The empirical response of corporate tax revenues is adjusted to eliminate the effect of profit repatriation. The results from the model are robust to changes in the intertemporal elasticity of substitution $\sigma$, the capital share $\alpha$, and to the introduction of additional deductions presents in the corporate tax code.}
\label{Theory_TCJA17_Qualitative}
\end{figure}

Since the empirical evidence on the TCJA-17 is not directly targeted, the exercise should be thought of as an out-of-sample forecasting exercise. The results from the model are presented and compared to the empirical evidence in \autoref{Theory_TCJA17_Qualitative}. The first column describes the response estimated in the data, and the second column the response from the model. The first row focuses on macroeconomic aggregates, and the second on c-corporate ones.

The model successfully forecasts the relative responses of aggregate and c-corporate variables estimated in the data. At the aggregate level, the model predicts a small response of output and investment, and a large fall in corporate tax revenues. Moreover, the response of investment is larger than that of output. At the c-corporate level, the model predicts an increase in payouts to shareholders larger than investment - in line with the data. Again, the response of investment is larger than that of output.

The intuition behind what happens can be broken down into two pieces. The first piece clarifies the response of c-corporations. Because of highly accelerated tax depreciation policy before the TCJA-17, the pre-reform corporate tax wedge was close to one ($\approx 0.97$ under the proposed calibration). As a result, the ability of the reform to further remove distortions was very limited in the first place, and  ended up providing little stimulus to c-corporate investment. At the same time, the tax-savings due to the reform were large, and c-corporations found themselves with a sizable amount of additional cash. Given their limited desire to increase investment, they distributed  a big share of this extra cash to their shareholders.

The second piece of intuition helps understand the even smaller response at the aggregate level. On the one hand, given a large share of pass-through businesses, the corporate provisions in the TCJA-17 applied to only  $60\%$ of the productive sector (measured in terms of economic activity). On the other hand, the remaining $40\%$   was not only not stimulated, but was in fact put at a competitive disadvantage relative to prior the reform, which produced a shift of economic activity from pass-through businesses to c-corporations. Overall, this resulted in  further dilution of the  aggregate stimulus.

\subsubsection{Endogenous Labor, CES Preferences, Variable Capital Utilization}
\label{Theory_Improving_Fit}
\begin{figure}[H]
\centerline{\includegraphics[scale=0.55]{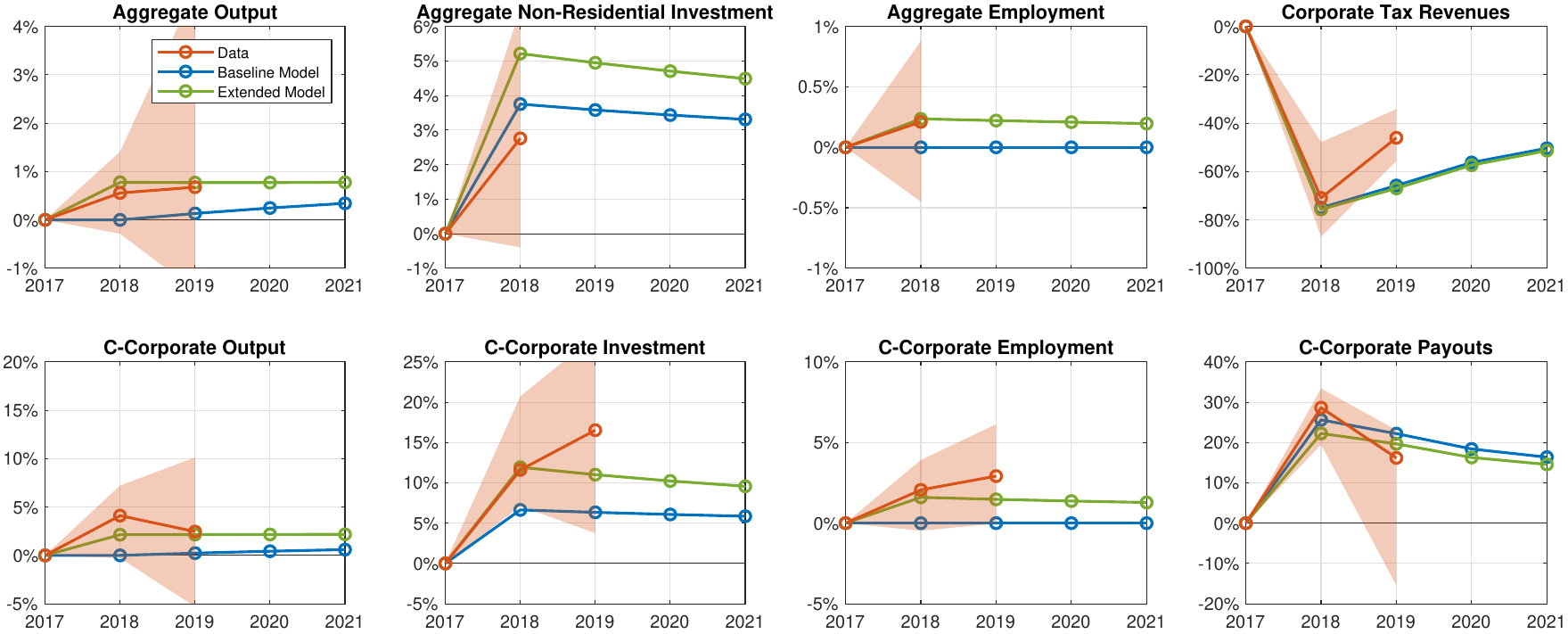}}
\caption{Quantitative Fit of the Baseline and Extended Model}
\label{Theory_TCJA17_Quantitative}
\end{figure}

The baseline model can forecast the overall pattern of macroeconomic and c-corporate responses, but is not able to offer a good quantitative fit for the response of some of the variables. In particular,  \autoref{Theory_TCJA17_Quantitative} shows that the response of output and investment for c-corporations is smaller than in the data. This is partly due to the assumption of exogenous labor supply - which reduces the ability of c-corporations to respond to the stimulus by hiring more workers - and partly due to inelastic capital supply in the short-term.\\

To assess the robustness of the proposed mechanism, I alter the baseline neoclassical model in three ways. First,  I endogenize labor supply and assume it is mobile across the two sectors. Second, I assume a more general CES consumption aggregator for the representative household. Third, I assume variable capital utilization. The additional  parameters are calibrated in a standard way and the details can be found in \autoref{Appendix_Extended_Model}. The ``extended model'' response is given by the green lines in \autoref{Theory_TCJA17_Quantitative}.

Endogenous labor supply that can move across the two sectors facilitates re-allocation of economic activity across sectors. Similarly, a CES consumption bundle allows  household's spending  to shift towards the goods produced by c-corporations - which are now relatively cheaper. Finally, variable capital utilization amplifies the response of c-corporate output as  it gives an additional margin of adjustment to the c-corporate sector.

Variable capital utilization interacts with corporate taxation in an interesting way. Since higher capital utilization accelerates the economic depreciation of capital, firms trade-off the marginal benefit of higher production with the marginal cost of replenishing the capital stock. By reducing the cost of capital, the TCJA-17 incentivizes higher capital utilization. \cite{ottonello2014capital} documents a large counter-cyclical share of idle productive capital, which is consistent with the proposed variable capital utilization mechanism.

\subsubsection{Decomposing the TCJA-17: Tax Rate Cut vs Bonus Depreciation}
\label{Theory_Rate_vs_Depreciation}
\begin{figure}[H]
\centerline{\includegraphics[scale=0.65]{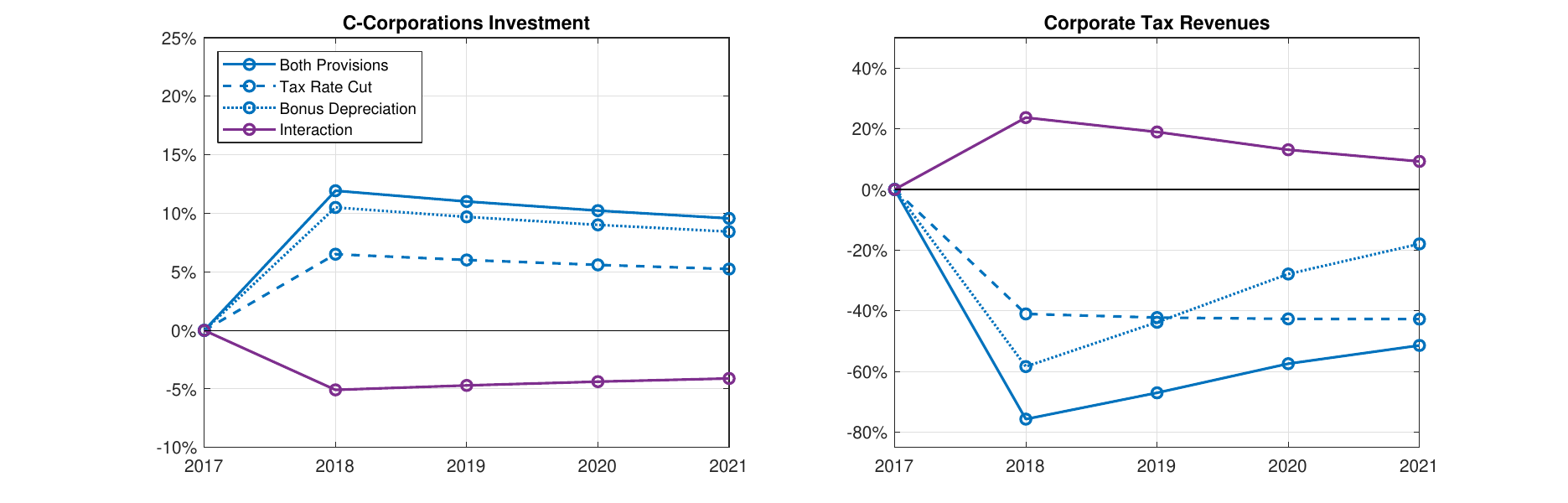}}
\caption{Decomposing the TCJA-17: Tax Rate Cut vs Bonus Depreciation}
\label{Theory_TCJA17_Rate_BonusDepreciation}
\end{figure}

I use the ``extended model'' to perform a counterfactual assessment of the importance of each of the two main corporate provisions in the TCJA-17, and the results for c-corporate investment and corporate tax revenues are reported in \autoref{Theory_TCJA17_Rate_BonusDepreciation}.

First, the expansionary effect of each provision on the investment of c-corporations is similar, as both are aimed at removing distortions to the investment decision.

Second, the interaction between these two provisions is negative. A cut to the corporate tax rate is more expansionary when the present discounted value of the tax depreciation schedule is lower. Similarly,  the effect of bonus depreciation is larger when the tax rate is higher. By reducing the tax rate while accelerating the depreciation schedule, the two provisions partially offset each other.

Third, the effect of these two provisions on corporate tax revenues is similar on impact, but is different in the long-run. A reduction of the tax rate produces a permanent loss of corporate tax revenues. An acceleration of the tax depreciation schedule, instead, results in a transitory one.

\subsection{TCJA-17 vs  Kennedy's Tax Cuts: Tax Multipliers}
This section compares the recent Trump's TCJA-17 with the  Kennedy's corporate tax cuts of the early 1960s through the lens of the theoretical framework proposed in the previous section. I use the ``extended model'' to assess the effects of each reform on GDP, aggregate investment and payouts to shareholders. By construction, the counterfactual experiment explains different macroeconomic outcomes through pre-existing differences in the corporate tax code, in the size of  the pass-through sector, and in the composition of the policy intervention. As a result, the exercise abstracts from differences in the economic environment - such as changes to market structure and technological change - and focuses on the differential effects caused by the tax code and the pass-through sector.

The TCJA-17 is simulated in the same way as before. Kennedy's corporate tax cuts are simulated as follow. I start from the calibration for 2017 and adjust the corporate tax rate, the tax depreciation rate, and the weights of the CES consumption aggregator to match corporate tax policy and the pass-through share in 1961. I then simulate Kennedy's tax cuts as unanticipated permanent changes to the following policy parameters:
\begin{itemize}[noitemsep]
  \item A permanent reduction in the corporate tax rate $\tau^{\pi}$ from $52\%$ to $48\%$.
  \item A permanent increase in the tax depreciation rate  $\delta^{\pi}$ from $0.10$ to $0.1857$.
\end{itemize}
As for the TCJA-17, the new tax depreciation rate applies only to new investment and further details can be found in \autoref{Appendix_New_Investment}.

\begin{figure}[H]
\centerline{\includegraphics[scale=0.70]{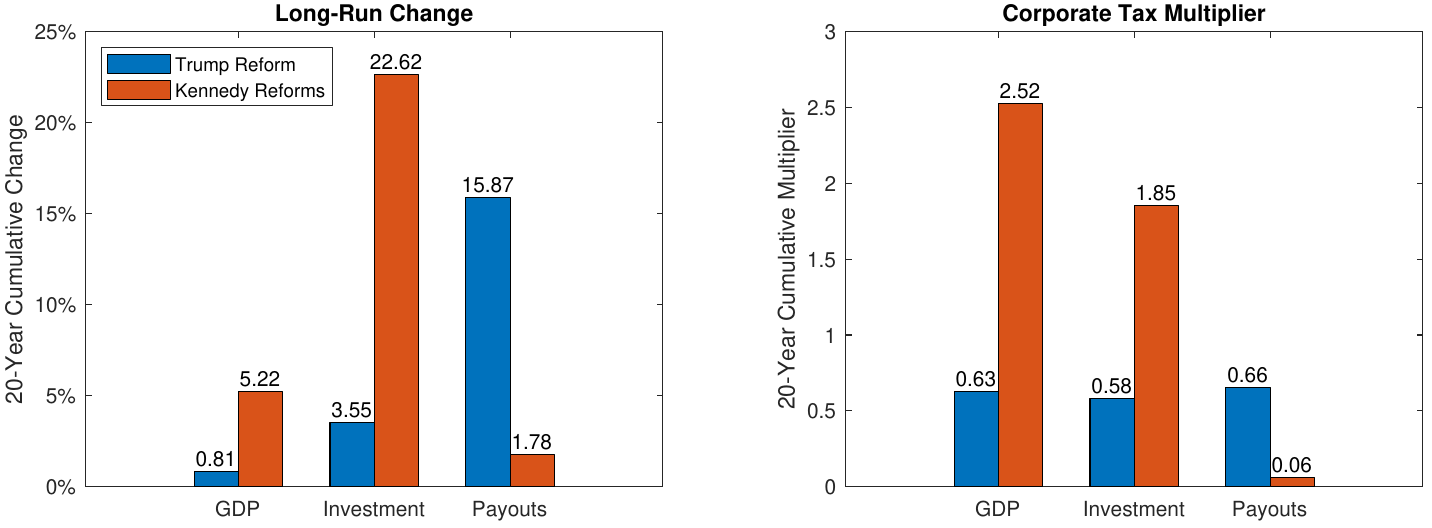}}
\caption{The TCJA-17 vs Kennedy's Corporate Tax Cuts \\ \scriptsize \textbf{Notes}: The long-run change is computed as the 20-year cumulative deviation from the steady-state, obtained by summing the level of each variable for 20 years after the reform and dividing it by its counterpart in the absence of the reform. The corporate tax multiplier is computed as the cumulative change in the level of each variable and divided by the cumulative change in corporate tax revenues.}
\label{Figure_Kennedy_Assessment}
\end{figure}

The results are reported in \autoref{Figure_Kennedy_Assessment}. In response to Kennedy's tax cuts, the model predicts a large increase in GDP and investment, and a small effect on payouts to shareholders: the opposite of Trump's Tax Cuts and Jobs Act. Similarly, the corporate tax multiplier for Kennedy's tax cuts is around $2.5$ for GDP, $1.85$ for investment, and close to zero for payouts to shareholders. For the TCJA-17, the multiplier is around $0.6$ for each variable. For every dollar of lost corporate tax revenues, Kennedy's corporate tax cuts stimulated GDP four times more than the TCJA-17.

The intuition behind these results is the following. In the early 1960s, the corporate tax rate was high and tax depreciation policy was not accelerated as it was mimicking economic depreciation. As a result, the corporate tax wedge was well below one (around $0.72$) before the reform. Kennedy's tax cuts increased the wedge significantly (to around $0.84$), thus providing strong stimulus to the investment of c-corporations. Moreover, since around $75\%$ of economic activity was taking place in the c-corporate sector, the aggregate effect was less diluted than in 2017.

\begin{figure}[H]
\centerline{\includegraphics[scale=0.70]{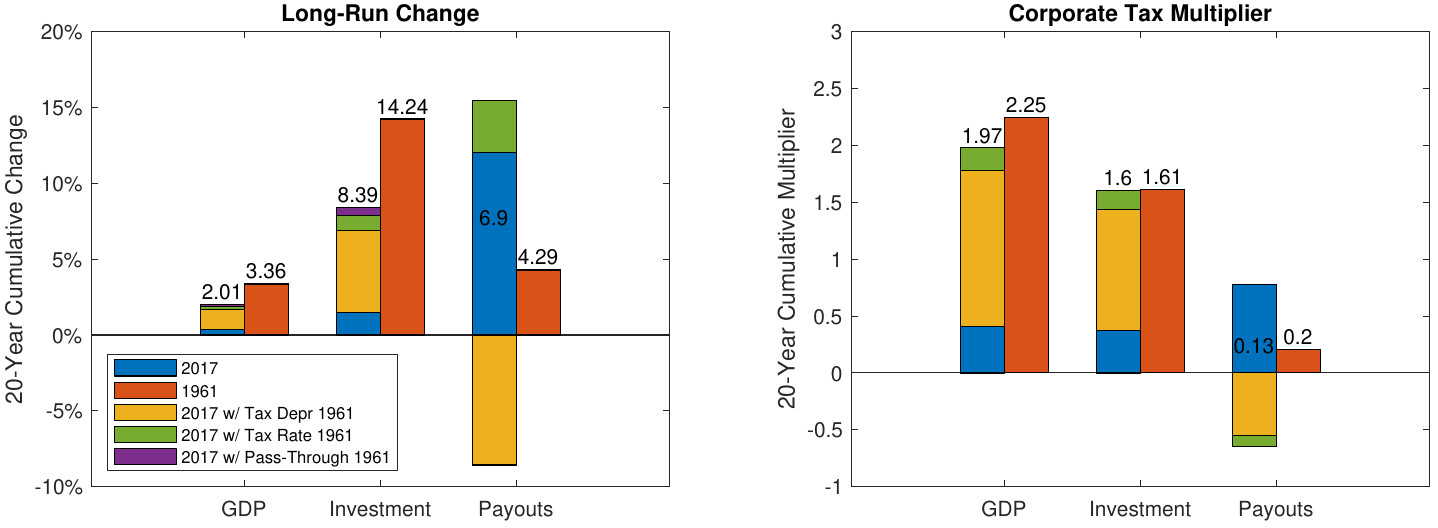}}
\caption{Understanding the Difference between the TCJA-17 and  Kennedy's Reforms}
\label{Figure_Kennedy_Decomposition}
\end{figure}

To better understand how each factor (i.e. tax rate, tax depreciation, pass-through share, policy intervention) contributed to the outcomes reported in \autoref{Figure_Kennedy_Assessment}, I perform another counterfactual experiment.  First, I control for differences in policy interventions by simulating the exact same reform in both 1961 and 2017: an unanticipated permanent reduction in the corporate tax rate by $10\%$. Then, I start from the calibration for 2017 and simulate the reform after changing one of the tax rate, tax depreciation rate and pass-through share at a time. So, for example, I take the calibration for 2017, set the tax depreciation rate equal to that in 1961, and simulate the reform. I repeat the same for the tax rate and the pass-through share. The results are reported in \autoref{Figure_Kennedy_Decomposition}.

The exercise shows that differences in tax depreciation policy between the early 1960s and 2017 account for most of the difference in the macroeconomic response to the reform. Looking at long-run changes, differences in pre-reform corporate tax rates and in the pre-reform share of pass-through businesses contribute similarly to the difference between the two reforms. The interaction between these three factors, instead, can be assessed by looking at the difference between the first and the second vertical bar for each variable. For example, under the 1961 calibration, the long-run investment response is $+14.24\%$, while the response under the 2017 calibration with each factor introduced at a time is only $+8.39\%$, which implies an interaction effect of $+5.85\%$.

The corporate tax multiplier features smaller interaction effects and seems unaffected by the size of the pass-through sector. This happens because a smaller pass-through sector implies  larger aggregate stimulus after a corporate tax cut, but also a larger loss of corporate tax revenues - since a larger share of the economy receives the tax cut. These two effects almost perfectly offset each other in this specific experiment.

\section{Relation to the Macroeconomics Literature}
This section relates the results in this paper to the broader macroeconomics literature. Since it is common in macroeconomics to think about corporate taxation as a form of capital taxation, I first provide a formal mapping between capital taxes and corporate taxes. Corporate taxes and capital taxes are simply two different tax instruments, and this explains why a large corporate tax reduction may fail to provide a large stimulus to production and capital accumulation.

Second, I show how common macroeconomics approaches to model corporate taxes compare when trying to predict the effects of the TCJA-17. I show that abstracting from either pass-through businesses or tax depreciation policy generates a response of investment and of payouts to shareholders that is inconsistent with the data.

\subsection{Corporate Taxes vs Capital Taxes}
\label{Analytic_Capital_Corporate}
It is insightful to relate the corporate tax proposed in this paper with the familiar concept of a ``capital income tax'', i.e. a tax imposed on the income produced by the productive factor ``capital''. Under a constant return-to-scale (CRS) technology, it is possible to unambiguously define capital income using Euler Theorem. Aggregate output can be expressed as:
\begin{align*}
  \hat{Y}_t =& Y_t + p_t \tilde{Y}_t \\
  =&  MPK_t k_t + MPL_t  l_t + p_t \cdot \Big( \tilde{MPK}_t \tilde{k}_t + \tilde{MPL}_t  \tilde{l}_t \Big) \\
  =&\underbrace{MPK_t k_t + p_t \tilde{MPK}_t \tilde{k}_t}_{\text{capital income}} + \underbrace{MPL_t  l_t + p_t \tilde{MPL}_t  \tilde{l}_t}_{\text{labor income}}.
\end{align*}
The best way to compare corporate taxes and capital taxes is to compare their tax bases. The corporate tax base is given by:
\[ TB^{\pi}_t = Y_t - w_t l_t - ID_t^{\pi}. \]
With a competitive labor market one has that $w_t = MPL_t$, and by Euler Theorem $Y_t - MPL_t l_t = MPK_t k_t$. As a result, the corporate tax base can be expressed as
\[ TB^{\pi}_t = \begin{cases} MPK_t k_t & \text{if} \ \delta^{\pi} = 0 \\
 MPK_t k_t -  \sum_{j=0}^{+\infty}  \textcolor{black}{\delta^{\pi}  (1 - \delta^{\pi})^j}  i_{t-j} & \text{if} \ \delta^{\pi} \in (0,1) \\
 MPK_t k_t - i_t & \text{if} \ \delta^{\pi} = 1.
\end{cases} \]
In other words, corporate taxes are levied on the capital income produced by the c-corporate sector reduced by a deduction for present (and past) investment.

The capital income tax base ($TB^K_t$), instead, is given by
\[ TB_t^K = MPK_t k_t + p_t \tilde{MPK}_t \tilde{k}_t. \]

Notice that - as long as production is CRS and factor markets are competitive -  capital taxes and corporate taxes are simply two different taxes levied on two different tax bases. The introduction of capital taxes in the model is more intuitive when the household accumulates the capital stock, and the introduction of corporate taxes is more intuitive when the productive sector accumulates the capital stock. Nonetheless, both can be introduced in the same economic environment following the approach above.  To better see this point,  \autoref{Appendix_Equivalent_Decentralization} explicitly introduces capital and corporate taxes in the context of the baseline model of \autoref{Theory_Baseline_Model}.

The comparison between the corporate tax base and the capital tax base reveals two main points. First, corporate taxes feature an investment deduction shaped by tax depreciation policy that is absent for capital taxes. Second, capital taxes apply to the income generated by the productive capital in all sectors of the economy, including pass-through businesses, while corporate taxes are levied only on c-corporations.

When there is no pass-through sector in the economy (i.e. when $\gamma = 1$), the difference between capital taxes and corporate taxes boils down to the investment deduction.

\subsubsection{Corporate Tax Revenues Collection}
Under full-expensing of investment (i.e. $\delta^{\pi} = 1$) in a frictionless environment, the corporate tax wedge becomes one and the distortion to the Euler Equation for capital accumulation disappears. It is interesting to see whether the corporate tax can actually collect revenues in such a case. In light of the relation between corporate and capital taxes, this is fundamentally the same question asked in \cite{abel2007optimal}.

\begin{prop}
    \textit{Capital tax revenues collection is positive. Corporate tax revenues collection can be positive, zero, or negative. More precisely:}
    \[ sign(T_t^{\pi}) = \begin{cases}     > 0  &\text{if} \quad \frac{i_t}{Y_t} < \alpha \\     = 0 &\text{if} \quad  \frac{i_t}{Y_t} = \alpha \\ <0  &\text{if} \quad   \frac{i_t}{Y_t} > \alpha  \end{cases}     \]
\end{prop}

\begin{proof}
See \autoref{Appendix_Corporate_Revenues_1}
\end{proof}

\begin{prop}
    \textit{A corporate tax with full-expensing of investment collects positive tax revenues in steady-state, i.e. $T^{\pi}_{ss} > 0$.}
\end{prop}

\begin{proof}
See \autoref{Appendix_Corporate_Revenues_2}
\end{proof}

\subsection{Alternative Approaches to Model Corporate Taxes}
Existing research in macroeconomics does not  explicitly model tax depreciation policy and pass-through businesses at the same time. In this section, I assess how alternative ways of modeling the corporate tax code compare when simulating the TCJA-17.

The most common approach is to ignore the pass-through sector and set tax depreciation equal to economic depreciation. Other two common approaches are to consider tax depreciation policy but to the ignore pass-through sector, or to consider the pass-through sector but restricting tax depreciation to  economic depreciation.

\begin{figure}[H]
\centerline{\includegraphics[scale=0.85]{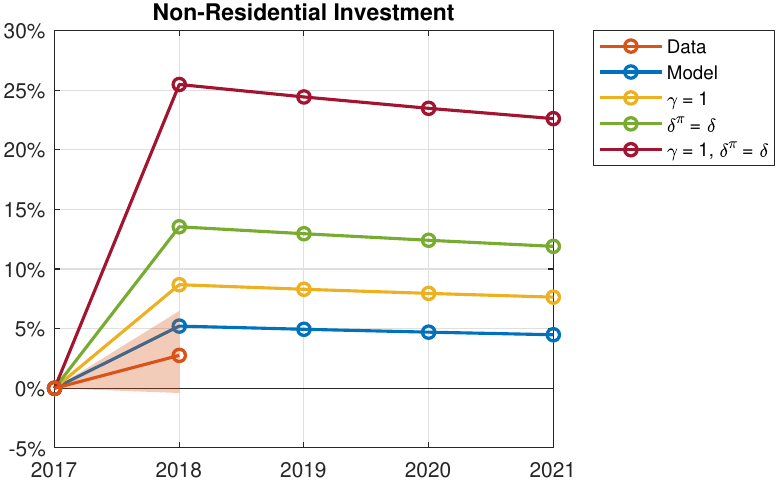}}
\caption{Investment Response to the TCJA-17: Alternative Modeling Approaches}
\label{Theory_TCJA17_Methodologies_Main}
\end{figure}

I simulate all three using the ``extended model'', and \autoref{Theory_TCJA17_Methodologies_Main} summarizes the response of aggregate business  investment.\footnote{Whenever tax depreciation is restricted to economic depreciation (i.e. $\delta^{\pi} = \delta$), I simulate the TCJA-17 by only reducing the corporate tax rate.} The yellow line summarizes the response of the economy shutting down the pass-through sector by assuming $\gamma = 1$. The green line restricts tax depreciation to economic depreciation by setting $\delta^{\pi} = \delta$. The red line imposes both $\gamma = 1$ and $\delta^{\pi} = \delta$.

Abstracting from the pass-through sector (yellow line) assumes that the entire productive sector benefits from the  corporate tax reduction, and  increases the response of investment in the model by roughly $70\%$. Retaining the pass-through sector but forcing tax depreciation to equal economic depreciation (green line) overestimates the response of investment by a factor of $3$. If $\delta^{\pi} = \delta$, the present discounted value of the tax depreciation schedule is low, and the corporate tax wedge introduces significant distortions to the investment decision of c-corporations. In this case, a corporate tax rate reduction provides sizable stimulus to investment. Finally, ignoring the pass-through sector and imposing economic depreciation (red line) overestimates the response of investment by a factor of $5$.

Restricting tax depreciation to economic depreciation corresponds to a very popular way to model the corporate tax base in macroeconomics. The corporate tax  is usually modeled as
\[ TB^{\pi}_t = Y_t - w_t l_t - \delta k_t \]
where $\delta k_t$ is a deduction for `capital depreciation'. In fact, this corresponds to a very specific tax depreciation policy. To see why, solve backwards the law of motion of capital accumulation
\[ \delta k_t = \delta i_{t-1} + \delta (1 - \delta) i_{t-2} + \delta (1 - \delta)^2 i_{t-3} + \dots \]
and plug it back into the corporate tax base
\[ TB^{\pi}_t = Y_t - w_t l_t - \sum_{j=1}^{+\infty}  \textcolor{black}{\delta (1 - \delta)^{j-1}}  i_{t-j}  \]
to see that tax depreciation equals economic depreciation with the caveat that the investment deduction can only be claimed one period after investment takes place.

\subsubsection{The Marginal Effective Tax Rate Approach}

\begin{figure}[H]
\centerline{\includegraphics[scale=0.85]{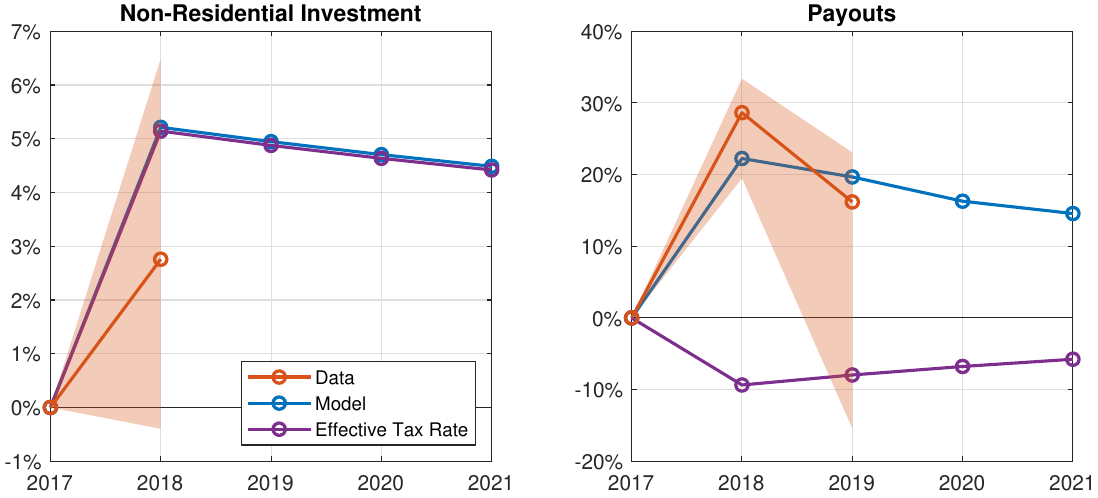}}
\caption{Comparison with the Effective Tax Rate Approach}
\label{Theory_TCJA17_Methodologies_Effective}
\end{figure}

A final alternative to model the corporate tax code is to abstract from tax depreciation policy and calibrate the corporate tax rate to a marginal effective tax rate which takes into account tax depreciation. I will label this as the ``effective tax rate approach'', and it is followed for example by \cite{acemoglu2020does}. To implement it in my framework, I set $\delta^{\pi} = 0$ and $\tau^{\pi} = \tau^*$, where $\tau^*$ is the marginal effective tax rate which is computed as follows
\[ \tau^* = 1 - \frac{1 - \tau^{\pi}}{1 - \lambda^{\pi} \tau^{\pi}}. \]
By construction, this marginal effective tax rate summarizes the ``corporate tax wedge'', and thus takes into account the effect of both tax depreciation policy and tax rate policy on the investment decision. Under my calibration, the marginal effective tax rates goes from $3.17\%$ before the TCJA-17 down to $0.34\%$ after it. The results are reported in \autoref{Theory_TCJA17_Methodologies_Effective}.

The effective tax rate approach produces a response of investment that is almost identical to the one obtained when modeling tax depreciation policy explicitly. This is not surprising, since the effective tax rate summarizes the corporate tax wedge. However, the effective tax rate cannot correctly model the level of corporate tax revenues and, as a result, fails to anticipate the response of payouts to shareholders observed in the data.

\section{Corporate Tax Policy over Time}
This section leverages the deterministic steady-state of the model to analyze the evolution of corporate tax policy in the US over the last few decades. To obtain these results, I start from the ``baseline model'' and assume away pass-through businesses (i.e.  $\gamma = l = 1$ ) and individual income taxes ($\tau^{II}=0$). Under these restrictions, I recover a neoclassical growth model featuring a corporate tax levied on the entire productive sector.

It is then possible to solve analytically for the deterministic steady-state of the model and express it as a function of its `undistorted' counterpart, i.e. the deterministic steady-state in the absence of corporate taxation (i.e. when $\tau^{\pi} = 0$). Long-run output ($Y_{ss}$) can be expressed as
\begin{gather*}
   Y_{ss} = Y_{ss}^* \cdot \omega_{ss}^{\frac{\alpha}{1-\alpha}} \\
   \text{where} \quad \omega_{ss} = \frac{1 - \tau^{\pi}}{1 - \lambda_{ss}^{\pi} \tau^{\pi}} \quad \text{and} \quad \lambda^{\pi}_{ss} = \frac{\delta^{\pi} (1 + \rho)}{\rho + \delta^{\pi}}
 \end{gather*}
Undistorted long-run output is given by $Y_{ss}^*$, and $\omega_{ss}$ is the corporate tax wedge in steady-state. Notice that $\lambda_{ss}^{\pi}$ is the present discounted value of the tax depreciation schedule in steady-state.

 In this frictionless environment, distortions to production are summarized by the corporate tax wedge - properly adjusted for capital intensity $\alpha$. Interestingly, corporate tax revenues and payouts to shareholders depend on the tax code in a more complicated way:
\begin{align*}
	T^{\pi}_{ss} &= \pi^*_{ss} \cdot \tau^{\pi} \cdot \Big[ \omega^{\frac{\alpha}{1 - \alpha}}  \cdot (1 + \frac{\delta}{\rho} \cdot (1 - \omega) ) \Big]	\\
	d_{ss} &= \pi^*_{ss} \cdot (1 - \tau^{\pi}) \cdot \Big[ \omega^{\frac{\alpha}{1 - \alpha}}  \cdot (1 + \frac{\delta}{\rho} \cdot (1 - \omega) )	\Big]
\end{align*}
and this duality further clarifies that the corporate tax code can differentially affect incentives and cash-flows, in line with what pointed out in \autoref{Theory_Wedge} and \autoref{Analytic_Capital_Corporate}.

\begin{figure}[H]
\centerline{\includegraphics[scale=0.75]{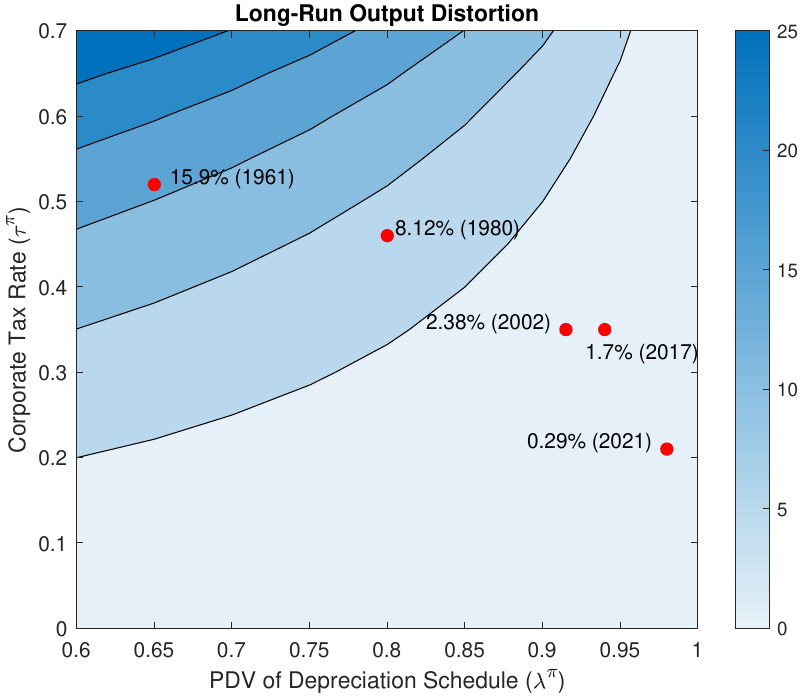}}
\caption{Corporate Tax Distortions over Time \\ \scriptsize \textbf{Notes}: Values for 1961 and 1980 are computed from \cite{cummins1994reconsideration}. Values for 2002, 2017 and 2021 are computed from \cite{zwick2017}. The only two parameters used  are $\beta = 0.94$ and $\alpha = 0.35$.}
\label{Figure_Analytic_Distortions}
\end{figure}

For convenience,  I then define the following measure of long-run distortions to output
\[ \text{Distortion}_{ss} =  1 - \frac{Y_{ss}}{Y_{ss}^*}  \]
and represent it in the corporate tax policy space in \autoref{Figure_Analytic_Distortions}. The figure displays a contour map of `isodistortions' for each combination of the corporate tax rate ($\tau^{\pi}$) and the present discounted value of the tax depreciation schedule ($\lambda_{ss}^{\pi}$). Red dots representing the corporate tax code in different years are superimposed to assess the evolution of corporate tax distortions over time.

The spirit of the exercise is to assess the level of distortions to GDP introduced by the corporate tax code using the deterministic steady-state of the model. The figure reveals a steady elimination of distortions by US policy-makers over time, captured by the movement towards the south-east corner of the map. For example, output was roughly $16\%$ lower than its undistorted counterpart before Kennedy's tax cuts, but only $1.7\%$ lower before the TCJA-17. This improvement have been achieved through several rounds of statutory tax rate cuts, changes to  tax depreciation rules, and repeated use of bonus depreciation over the decades.

While the numbers reported in the figure should be taken with a grain of salt, they teach two important lessons. On the one hand, corporate tax policy has become less distortionary over time. On the other hand, policy-makers are now running short of ammunition. Given that the current level of distortions is almost zero, further reductions of the statutory corporate tax rate and/or acceleration of the tax depreciation schedule will produce little stimulus to the US economy.

\section{Conclusions}
This paper has focused on tax depreciation policy and the distinction between c-corporations and pass-through businesses to understand the effects of two major corporate tax reforms in the US. While the proposed theoretical framework is intendedly stylized in order to make the transmission mechanism as transparent and robust as possible, it can be  enriched along several dimensions. For example, two candidates are the introduction of sectoral heterogeneity and the analysis of corporate debt and of the interest-payment deduction. Preliminary results suggest that these two extensions do not alter the overall predictions of the model, but they do allow the theory to generate additional implications for different sectors or for corporate leverage. This could be of interest on its own, or could be used to discipline the theory further by exploiting empirical evidence from the cross-section of firms or of industries.

\bibliographystyle{econ}
\bibliography{references}
\appendix
\setcounter{table}{0}
\setcounter{figure}{0}
\setcounter{footnote}{0}
\setcounter{page}{0}
\renewcommand{\thetable}{A\arabic{table}}
\renewcommand\thefigure{A\arabic{figure}}
\pagenumbering{arabic}
\renewcommand*{\thepage}{A\arabic{page}}
\section*{\LARGE Appendix}

\section{Empirics Details}

\subsection{TCJA-17 and Anticipation Effects}
\label{Appendix_Empirics_TCJA17_Anticipation}

One issue with this approach is that the TCJA-17 is arguably not the only shock hitting the US economy in this period. To assess the potential impact of unforeseen shocks, I compute historical forecast errors and use them to construct confidence intervals for the forecasts. These forecast intervals are directly informative about the errors made by forecasters in the past and, to the extent that these errors reflect unanticipated shocks, the intervals do as well. 

\begin{figure}[H]
\centerline{\includegraphics[scale=0.75]{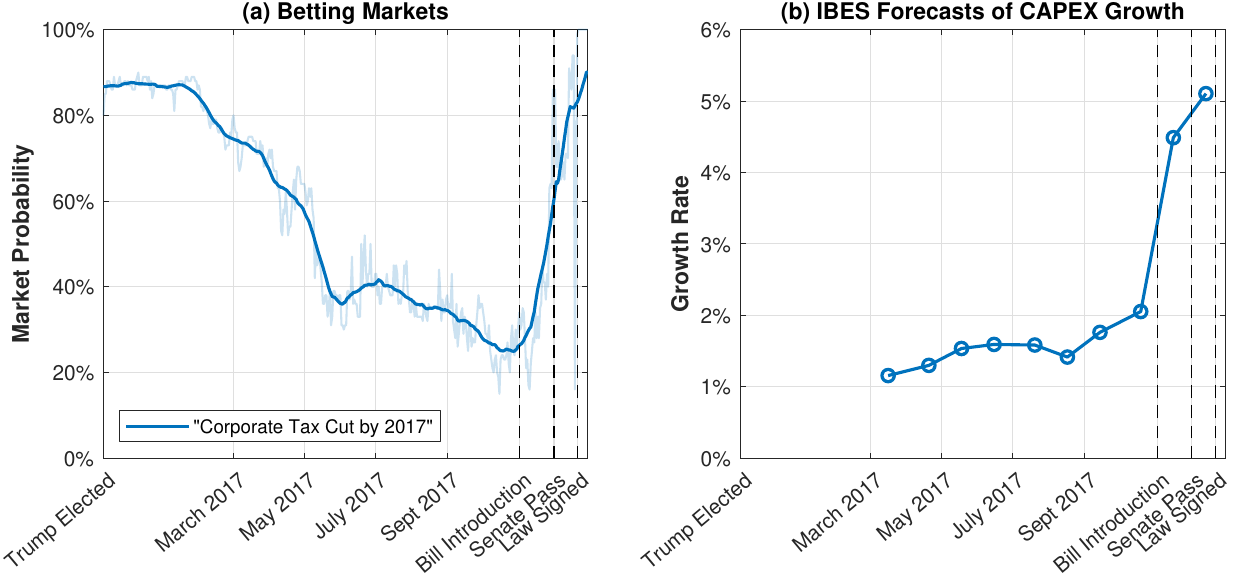}}
\caption{\footnotesize Perceptions of a Corporate Tax Reform before the TCJA-17 \\ \textbf{Source}: IBES, PredictIt.}
\label{Figure_TCJA17_Probability}
\end{figure}

Another important concern  is that pre-reform forecasts might incorporate expectations of an imminent reform, thereby biasing the estimated effect. To assess the extent of anticipation effects in pre-reform forecasts, I first look at the probability of an imminent reform from betting markets data. Panel (a) of  \autoref{Figure_TCJA17_Probability} reports the probability of a corporate tax cut from the election of former President Trump to the passage of the TCJA-17.

A corporate tax cut was perceived as almost certain in the first few months after the election, arguably as a reflection of electoral campaign promises. However, as months went by without any legislative action, the perceived probability decreased to around $30\%$ in the summer of 2017. It then picked up  once the first draft of the TCJA-17 reform bill was introduced into Congress in the Fall of 2017, and increased quickly as the bill passed congressional vote and eventually became law in December 2017.

Based on the  probability from betting markets, it appears that forecasts made in the summer of 2017 are the least likely to incorporate anticipation effects. It is possible, however, that  betting market participants' beliefs differ systematically  from those of professional forecasters. To mitigate this concern, I examine the dynamic evolution of professional forecasts from IBES in Panel (b) of \autoref{Figure_TCJA17_Probability}. The plot reports the  evolution over time of forecasts of capital expenditure growth for 2018. The series exhibits a strong correlation with betting market probabilities, which suggests a similar evolution of beliefs among betting market participants and professional forecasters.

While it is  not possible to completely rule out anticipation effects in pre-reform forecasts, it is important to realize that  there was no official draft of the reform before the Fall of 2017, and thus no clear indication of the magnitude and composition of a possible policy intervention. This consideration further mitigates concerns of anticipation effects.

\subsection{Dividends vs Share Repurchases around the TCJA-17}

\begin{figure}[H]
\centerline{\includegraphics[scale=0.75]{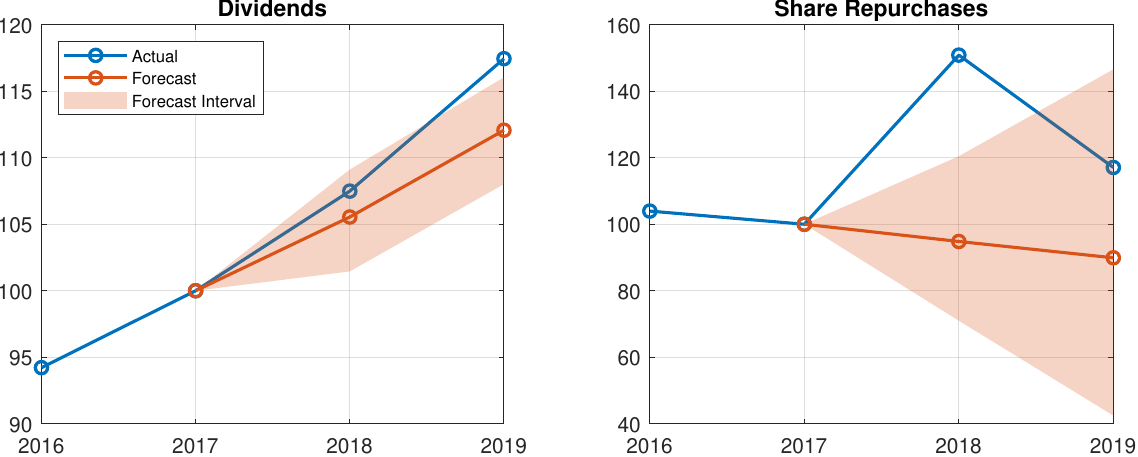}}
\caption{Decomposition of the Payouts Response to TCJA-17}
\label{TCJA17_Payouts_Decomposition}
\end{figure}

\subsection{Corporate Tax Revenues and Repatriated Earnings}
\label{Appendix_Empirics_Repatriation}

\begin{figure}[H]
\centerline{\includegraphics[scale=0.75]{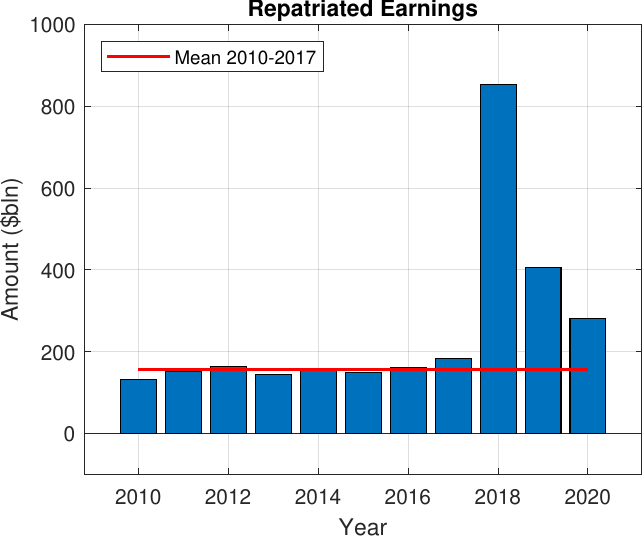}}
\caption{Repatriated Earnings during 2010-2020}
\label{TCJA17_Repatriated_Earnings}
\end{figure}

When comparing the model's predictions with the empirical evidence, I adjust corporate tax revenues to remove taxes paid on repatriated earnings. I follow \cite{smolyansky2019us} to measure earnings repatriated by corporations, which are reported in \autoref{TCJA17_Repatriated_Earnings}.

The adjustment to corporate tax revenues attempts to remove the revenues collected upon repatriation. This requires to first estimate repatriated earnings caused by the TCJA-17, and then to estimate the tax revenues collection upon them.

Let $RE_t$ be repatriated earnings at time $t$. I compute  repatriated earnings that exceed historical levels by subtracting the mean over the period 2010-2017 to estimate repatriated earnings caused by the TCJA-17:
\[ \hat{RE}_t = \begin{cases}
      0 & t\leq 2017 \\
      RE_t - \sum_{j = 2010}^{2017} RE_j & t>2017
   \end{cases} \]

\begin{figure}[H]
\centerline{\includegraphics[scale=0.75]{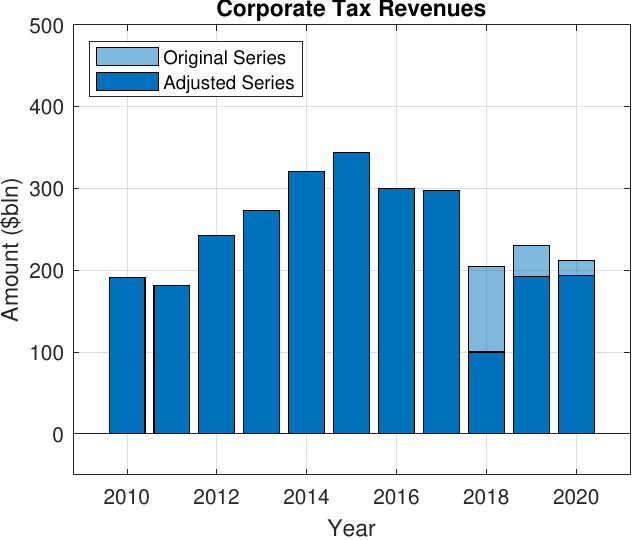}}
\caption{Corporate Tax Revenues during 2010-2020}
\label{TCJA17_Adjusted_Corporate_Revenues}
\end{figure}

The TCJA-17 introduced a repatriation tax on repatriated earnings of $15.5\%$ on cash and cash equivalents and of $8\%$ on earnings not held in cash or cash equivalents. Since I do not observe the composition of repatriated earnings, I assume that repatriation occurs mainly through cash and cash equivalents and assume a tax rate of $15\%$. Adjusted corporate-tax revenues are then computed as
\[ \tilde{T}^{\pi}_t = T_t^{\pi} - \hat{RE}_t. \]
The results are summarized in \autoref{TCJA17_Adjusted_Corporate_Revenues}.

\newpage
\section{Modeling Details}

\subsection{Calibration of the Tax Depreciation Schedule}
\label{Appendix_Calibration_Depreciation}

With both bonus depreciation and investment tax credits, the corporate tax liability is given by:
\[ T^{\pi}_t = \tau^{\pi} \cdot [Y_t - w_t l_t - ID_t^{\pi}] - c^{\pi} i_t, \quad \text{with } ID_t^{\pi} = b \cdot  i_t  + (1-b) \cdot \Delta^{\pi}(L) \cdot i_t\] 
where $b$ is the bonus depreciation rate, $c^{\pi}$ is the investment tax credit rate, and $\Delta^{\pi}(L)$ is the tax depreciation schedule represented as a lag-polynomial. It is immediate to see the convience of this representation:
\begin{align*}
\Delta^{\pi}(L) \cdot i_t &= (\delta^{\pi}_0 L^0 + \delta^{\pi}_1 L + \delta^{\pi}_2 L^2 + ... ) \cdot i_t \\
&= \delta^{\pi}_0 i_t + \delta^{\pi}_1 i_{t-1} + \delta^{\pi}_2 i_{t-2} + ...
\end{align*}
When we evaluate the tax depreciation schedule polynomial at the discount rate, we recover its present discounted value:
\[ \Delta^{\pi}(\beta) = \sum_{j=0}^{+\infty} \beta^{j} \delta^{\pi}_j \]
Thanks to this, one can immediately compute the present discounted value of tax depreciation policy and investment tax credits as:
\[ b + (1 - b) \cdot \Delta^{\pi}(\beta) + \frac{c^{\pi}}{\tau^{\pi}}. \]
Instead of explicitly modeling bonus depreciation and investment tax credits, one can therefore define an new tax depreciation schedule $\tilde{\Delta}^{\pi}(L)$ such that its present discounted value coincides with the value above:
\[ \tilde{\Delta}^{\pi}(\beta) = b + (1 - b) \cdot \Delta^{\pi}(\beta) + \frac{c^{\pi}}{\tau^{\pi}}. \]
Additional details on modeling tax depreciation policy can be found in the second chapter of \cite{furno2022essays}. 

To calibrate an auxiliary tax depreciation schedule reflecting both bonus depreciation and investment tax credits, I select the policy parameter $\delta^{\pi}$ to match an estimate of the present discounted value (PDV) of the overall tax depreciation policy from the existing literature. Given a discount rate $\beta$, the present discount value of the auxiliary tax depreciation schedule as a function of $\delta^{\pi}$ is given by 
\[  \lambda^{\pi} =  \sum_{j=0}^{+\infty} \beta^j \cdot \delta^{\pi} \cdot (1 - \delta^{\pi})^j = \frac{\delta^{\pi}}{1 - \beta \cdot (1 - \delta^{\pi})} \]
and therefore 
\[ \delta^{\pi} = \frac{\rho \cdot \lambda^{\pi}}{1 + \rho - \lambda^{\pi}} \quad \text{where} \quad \rho \equiv \frac{1 - \beta}{\beta} \]

\subsubsection*{TCJA-17}

To calibrate $\delta^{\pi}$ in 2017, I build on \cite{zwick2017}. I start from their cross-sectoral average of the investment-weighted PDV of MACRS depreciation rules and add bonus depreciation: they estimate a present discounted value of deductions of $0.879$ to which I add an existing  $50\%$ bonus depreciation. The resulting present discounted value is given by 
  \[ 0.50 + (1 - 0.50) \times 0.879 = 0.9395 \]
and the associated  $\delta^{\pi}$ equals $0.4823$. 

To calibrate the new value of $\delta^{\pi}$ after the TCJA-17, I increase bonus depreciation from $50\%$ to $90\%$. As discussed in the main text, I do not increase bonus depreciation to $100\%$ to account for some restrictions on asset eligibility. This implies an increase in the present discounted value of tax depreciation policy from $0.9395$ to $0.9879$, and a new value of $\delta^{\pi} = 0.8305$.

\subsubsection*{Kennedy's Tax Cuts}
To calibrate tax depreciation policy before and after Kennedy's tax cuts I follow \cite{cummins1994reconsideration}. They estimate a present discounted value of the tax depreciation schedule for equipment of $0.6473$ in 1960.\footnote{In Table 1 of their paper, the present discounted value of depreciation is multiplied by the statutory tax rate. They report a value of $0.3366$ for 1960, which becomes $0.6473$ after dividing by a statutory tax rate equal to $0.52$.} Since this is almost the present discounted value under economic depreciation, I set $\delta^{\pi} = \delta = 0.10$.

For 1965, they estimate a present discounted value of the tax depreciation schedule equal to $0.7258$, which is accompanied by an investment tax credit equal to $0.0657$. Following the discussion above, I divide the investment tax credit by the statutory tax rate in 1965 and obtain a value of $0.1369$, which leads to an overall present discounted value of $0.8627$. This leads to a new value of $\delta^{\pi}$ equal to $0.2737$.

\subsection{Simulating Bonus Depreciation on New Investment}
\label{Appendix_New_Investment}
To capture the fact that bonus depreciation applies to new investment - as opposed to past investment not depreciated yet - I introduce auxiliary variables. Let $\delta^{\pi, B}$ and $k_t^{\pi, B}$ be the tax depreciation rate and the stock of un-depreciated investment before the reform.  Let $\delta^{\pi, A}$ and $k_t^{\pi, A}$ be the same variables after the reform. Finally, let $D^A_t$ take value equal to one after the reform and equal to zero before.\\
I then rewrite the investment deduction as follows
\begin{align*}
ID_t^{\pi} &= \delta^{\pi, B} \cdot \Big[ (1 - D^A_t) \cdot i_t + k_t^{\pi, B} \Big] + \delta^{\pi, A} \cdot \Big[ D^A_t \cdot i_t + k_t^{\pi, A} \Big]
\end{align*}
where
\begin{align*}
  k_{t+1}^{\pi, B} &= (1 - \delta^{\pi, B}) \cdot \Big[ (1 - D_t^A) \cdot i_t + k_t^{\pi, B}  \Big] \\
  k_{t+1}^{\pi, A} &= (1 - \delta^{\pi, A}) \cdot \Big[  D_t^A \cdot i_t + k_t^{\pi, A}  \Big]
\end{align*}
This modeling strategy ensures that - after the reform - past investment that has not been depreciated yet can still be depreciated using the old depreciation schedule, while new investment is depreciated using the new depreciation schedule.
\subsection{Extended Model Details}
\label{Appendix_Extended_Model}
The `extended model' starts from the `baseline model' and introduces: 1) endogeneous labor supply that is mobile across sectors; 2) a CES consumption aggregator; 3) variable capital utilization in the c-corporate sector.\\
The representative household solves the following optimization problem:
\begin{gather*}
\max \quad   \sum_{t=0}^{+\infty} \beta^t \Bigg[ \frac{\hat{c}_t^{1 - \sigma}}{1 - \sigma}  - \frac{\hat{l}_t^{1 + \phi}}{1 + \phi} \Bigg] \\
s.t. \quad 	\hat{c}_t = \Big( \eta \cdot c_t^{\epsilon} + (1 - \eta) \cdot \tilde{c}_t^{\epsilon} \Big)^{\frac{1}{\epsilon}} \\
        \hat{l}_t = l_t + \tilde{l}_t \\
			c_t + p_t \tilde{c}_t + \Delta S_{t+1} P_t + \Delta \tilde{S}_{t+1} \tilde{P}_t = (1 - \tau^{II}) \cdot \big[ w_t \hat{l}_t + S_t d_t + \tilde{S}_t \tilde{d}_t \big] + \text{Transfer}_t   \\
      \Lambda_{t+j,t} \equiv \beta^j \cdot \frac{u'(\hat{c}_{t+j})}{u'(\hat{c}_{t})} \cdot \frac{\partial \hat{c}_{t+j} / \partial c_{t+j}}{\partial \hat{c}_{t} / \partial c_t}
\end{gather*}

The productive sector solves the following optimization problems:
\begin{multicols}{2}
\textbf{C-Corporations}
  \begin{align*}
  \max \quad &  \sum_{t=0}^{+\infty} \Lambda_{0,t} d_t \\
  s.t. \quad 	d_t &=   \pi_t - T_t^{\pi}  \\
      \pi_t &= Y_t - w_t l_t - i_t\\
      k_{t+1} &= (1 - \delta(u_t)) \cdot k_t + i_t \\
      \delta(u_t) &= \delta_0 + \delta_1 (u_t - 1) + \frac{\delta_2}{2} (u_t - 1)^2 \\
      Y_t &=  (u_t \cdot k_t)^{\alpha} \cdot l_t^{1 - \alpha} \\
      \textcolor{black}{T_t^{\pi}} &\textcolor{black}{= \tau^{\pi} \cdot TB_t^{\pi}} \\
      \textcolor{black}{TB_t^{\pi}} &\textcolor{black}{= Y_t - w_t l_t - ID^{\pi}_t}
  \end{align*}

\columnbreak
\textbf{Pass-Through Businesses}
\begin{align*}
\max \quad &  \sum_{t=0}^{+\infty} \Lambda_{0,t} \tilde{d}_t \\
s.t. \quad 	\tilde{d}_t &=   \tilde{\pi}_t \\
    \tilde{\pi}_t &= p_t \cdot \Big( \tilde{Y}_t - w_t \tilde{l}_t -  \tilde{i}_t \Big)\\
    \tilde{k}_{t+1} &= (1 - \tilde{\delta}) \tilde{k}_t + \tilde{i}_t \\
    \tilde{Y}_t &=  \tilde{k}_t^{\tilde{\alpha}} \cdot l_t^{1 - \tilde{\alpha}}
\end{align*}
\end{multicols}

The government collects revenues and channels them into wasteful spending and transfers:
\begin{align*}
 T_t &= T^{\pi}_t + T^{II}_t \\
 G_t &= \theta \cdot T_t \\
 \text{Transfer}_t &= (1 - \theta) \cdot T_t
\end{align*}

The new parameters introduced are $\phi$, $\delta_0$, $\delta_1$, $\delta_2$,  $\eta$ and $\epsilon$. I set $\phi = 4$, which implies a Frisch elasticity of $0.25$. The steady-state economic depreciation for c-corporations is given by $\delta_0 = 0.10$ since I set $\delta_1 = \frac{1}{\beta} - (1 - \delta_0) = 0.1638$. The parameter $\delta_2$ is set equal to $0.10$ to target a steady-state elasticity of depreciation to utilization of approximately $0.60$, which is basically the mid-point between  the values in \cite{basu1997cyclical} and  \cite{king1999resuscitating}.\\
I set $\epsilon = 0.33$ to target an elasticity of substitution between the goods produced by the two sectors of approximately $1.5\%$. This implies some substitutability between the two varieties. Given $\epsilon$, I use $\eta$ to calibrate the target the relative size of c-corporations. I set $\eta = 0.55$ for the TCJA-17, and $\eta = 0.70$ for Kennedy's tax cuts.

\newpage
\subsection{An Equivalent Decentralization}
\label{Appendix_Equivalent_Decentralization}

\subsubsection{Capital Taxes in the Baseline Model}
To understand how to introduce capital taxes in the baseline model presented in \autoref{Theory_Baseline_Model}, start from the capital tax base and apply Euler Theorem:
\begin{align*}
TB_t^K =& MPK_t k_t + p_t \tilde{MPK}_t \tilde{k}_t \\
=& Y_t - w_t l_t + p_t \cdot \Big(\tilde{Y}_t - \tilde{w}_t \tilde{l}_t \Big).
\end{align*}
This implies that capital taxes can be levied by imposing a tax on operating income (i.e. revenues minus wages) on both the c-corporate sector and the pass-through sector.

\subsubsection{An Alternative Decentralization of the Baseline Model}
Since it is not immediate to see that a tax on the operating income of both sectors is equivalent to taxing capital income in the economy, I introduce explicitly capital taxes in an equivalent - but more familiar - decentralization of the model where the household accumulates the capital stock. For clarify of exposition and without loss of generality, let's assume away individual income taxation (i.e. $\tau^{II} = 0$). The representative household's problem is now given by

\begin{gather*}
\max \quad   \sum_{t=0}^{+\infty} \beta^t \frac{\hat{c}_t^{1 - \sigma}}{1 - \sigma}   \\
s.t. \quad 	\hat{c}_t = c_t^{\gamma} \cdot \tilde{c}_t^{1 - \gamma} \\
			c_t + p_t \tilde{c}_t + k_{t+1} + p_t \tilde{k}_{t+1} =  w_t l_t + p_t \tilde{w}_t \tilde{l}_t + k_t (1 - \delta + R_t) + p_t \tilde{k}_t (1 - \tilde{\delta} + \tilde{R}_t) + \pi_t  + \tilde{\pi}_t  - T_t^K + \text{Transfer}_t  \\
			l_t + \tilde{l}_t = 1, \quad l_t = l, \quad \tilde{l}_t = \tilde{l} \\
      k_{t+1} = (1 - \delta) k_t + i_t \\
      \tilde{k}_{t+1} = (1 - \tilde{\delta}) \tilde{k}_t + \tilde{i}_t \\
     T^K_t = \tau^{k} \cdot \Big(k_t R_t + p_t \tilde{k}_t \tilde{R}_t \Big)
\end{gather*}
The representative household supplies labor and rents capital to the productive sector, and the factor markets are competitive. The productive sector's problem is given by:

\newpage

\begin{multicols}{2}
\textbf{C-Corporations}
  \begin{align*}
  \max_{\{\pi_t, Y_t, k_t, l_t\}} \quad &  \sum_{t=0}^{+\infty} \Lambda_{0,t} \pi_t \\
  s.t. \quad 	\pi_t &= Y_t - w_t l_t - R_t k_t \\
      Y_t &=  k_t^{\alpha} \cdot l_t^{1 - \alpha}
  \end{align*}

\columnbreak
\textbf{Pass-Through Businesses}
\begin{align*}
\max_{\{\tilde{\pi}_t, \tilde{Y}_t, \tilde{k}_t, \tilde{l}_t\}} \quad &  \sum_{t=0}^{+\infty} \Lambda_{0,t} \tilde{\pi}_t \\
s.t. \quad 	\tilde{\pi}_t &= p_t \cdot \Big( \tilde{Y}_t - \tilde{w}_t \tilde{l}_t -  \tilde{R}_t \tilde{k}_t \Big)\\
    \tilde{Y}_t &=  \tilde{k}_t^{\tilde{\alpha}} \cdot l_t^{1 - \tilde{\alpha}}
\end{align*}
\end{multicols}

The government's behavior and aggregation are exactly as in \autoref{Theory_Baseline_Model}. This alternative decentralization is equivalent to the one in \autoref{Theory_Baseline_Model} in the sense that the equilibrium law of motion of the system is the same. Since the factor markets are competitive, we have that $R_t = MPK_t$ and $\tilde{R}_t = \tilde{MPK}_t$. As a result, in equilibrium we have that capital taxes are equal to
\[ T^K_t = \tau^{k} \cdot \Big(MPK_t k_t + p_t \tilde{k}_t \tilde{MPK}_t \Big) \]
and therefore the capital tax base is equal to
\[ TB_t^K = MPK_t k_t + p_t \tilde{MPK}_t \tilde{k}_t.  \]

It is also possible to introduce corporate taxes in this decentralization of the model. To do so, replace the capital tax $T^K_t$ with a corporate tax $T^{\pi}_t$ - still levied on the household - given by:
\[ T^{\pi}_t = \tau^{\pi} \cdot \Big(R_t k_t - \sum_{j=0}^{+\infty}  \textcolor{black}{\delta^{\pi}  (1 - \delta^{\pi})^j}  i_{t-j} \Big) \]
where $\delta^{\pi}$ is the tax depreciation rate. Given a competitive capital rental market  and a competitive labor market we have that $R_t = MPK_t$ and that $w_t = MPL_t$. By applying Euler Theorem we can express corporate taxes as:
\[ T^{\pi}_t = \tau^{\pi} \cdot \Big(Y_t - w_t l_t - \sum_{j=0}^{+\infty}  \textcolor{black}{\delta^{\pi}  (1 - \delta^{\pi})^j}  i_{t-j} \Big). \]

\newpage
\subsection{Proofs of Corporate Tax Revenues Collection}

\subsubsection{Proof of Proposition 1}
\label{Appendix_Corporate_Revenues_1}
Let's first consider a capital tax. Capital tax revenues are given by
\[ T^K_t = \tau^{k} \cdot \Big(MPK_t k_t + p_t \tilde{k}_t \tilde{MPK}_t \Big) \]
and, since $\tau^k > 0$, their sign is given by
\[ sign(T_t^k)  = sign(MPK_t k_t + p_t \tilde{k}_t \tilde{MPK}_t ). \]
In a non-degenerate (or interior) equilibrium, one has that $k_t > 0$, $MPK_t > 0$, $\tilde{k}_t >0$, $\tilde{MPK}_t > 0$, and $p_t > 0$. Therefore:
\[ sign(T_t^k) > 0. \]

Let's now consider a corporate tax with full-expensing of investment. Corporate tax revenues are given by
\[ T^{\pi}_t = \tau^{\pi} \cdot \Big(MPK_t k_t - i_t \Big) \]
and, since $\tau^{\pi} > 0$, their sign is given by
\[ sign(T_t^{\pi})  = sign(MPK_t k_t - i_t ). \]
Because of a Cobb-Douglas production function, one has that $MPK_t k_t = \alpha Y_t$. Therefore
\begin{align*}
  sign(T_t^{\pi})  =& sign(\alpha Y_t - i_t ) \\
  =& sign\Big(\alpha - \frac{i_t}{Y_t} \Big)
\end{align*}
where the second equality follows from the fact that $Y_t > 0$. If $\alpha > \frac{i_t}{Y_t}$, corporate tax revenues collection is positive; if  $\alpha < \frac{i_t}{Y_t}$, it is negative; if  $\alpha = \frac{i_t}{Y_t}$, it is zero.

\subsubsection{Proof of Proposition 2}
\label{Appendix_Corporate_Revenues_2}
Consider the corporate tax base in steady-state :
\[ TB_{ss}^{\pi} = Y_{ss} - w_{ss} l_{ss} - ID_{ss} \]
and because of full-expensing we have $ID_{ss} = i_{ss}$. Apply Euler Theorem to get
\[ TB_{ss}^{\pi} = MPK_{ss} \cdot k_{ss} - i_{ss}. \]
Since $i_{ss} = \delta k_{ss}$, the tax base can be rewritten as
 \[ TB_{ss}^{\pi} = k_{ss}  \cdot (MPK_{ss} - \delta) \]
which implies that  corporate tax revenues collection is positive if $MPK_{ss} - \delta > 0$. In steady-state, the Euler Equation for capital accumulation  becomes
\[ 1  = \beta \Big[1 - \delta + MPK_{ss} \Big] \]
which implies that
\[ MPK_{ss} - \delta = \frac{1}{\beta} - 1 \equiv \rho \]
where $\rho > 0$ is the rate of time preferences. Since $\rho >0$, we have that
\[ TB^{\pi}_{ss} > 0. \]

\end{document}